%% file: whitepaper_v1.tex
\documentclass[aps, prd, superscriptaddress,nofootinbib,twocolumn,preprintnumbers]{revtex4-1}

\usepackage{amsmath,amssymb,amsfonts,bm,slashed,graphicx,array,multirow,xspace}
\usepackage{placeins}
\usepackage{setspace}
\usepackage[labelfont=bf,font={small,stretch=1}, justification=centerlast]{caption}
\usepackage[shortlabels]{enumitem}

\newcommand{\mm}{\text{mm}}
\newcommand{\cm}{\text{cm}}

\newcommand{\CODEXb}{\mbox{CODEX-b}\xspace}
\newcommand{\CODEXbeta}{\mbox{CODEX-$\beta$}\xspace}

\allowdisplaybreaks

\makeatletter
\g@addto@macro\bfseries{\boldmath}
\makeatother

\usepackage[dvipsnames]{xcolor}

\usepackage{fancyhdr}
\definecolor{nicered}{rgb}{0.7,0.1,0.1}
\definecolor{nicegreen}{rgb}{0.1,0.5,0.1}
\definecolor{niceblue}{rgb}{0.1,0.1,0.7}
\usepackage[bookmarks=false]{hyperref}
\hypersetup{colorlinks,citecolor= nicegreen,linkcolor= niceblue}

\begin{document}

\title{The Road Ahead for CODEX-b\\
{\normalsize \normalfont A Snowmass whitepaper}}

\include{authors_full}


\begin{abstract}
In this Snowmass contribution we present a comprehensive status update on the progress and plans for the proposed CODEX-b detector,
 intended to search for long-lived particles beyond the Standard Model.
We review the physics case for the proposal and present recent progress on optimization strategies for the detector and shielding design, 
as well as the development of new fast and full simulation frameworks.
A summary of the technical design for a smaller demonstrator detector (\CODEXbeta) for the upcoming Run~3 of the LHC is also discussed, 
alongside the road towards realization of the full experiment at the High-Luminosity LHC.
\end{abstract}

\preprint{\includegraphics[width=3cm]{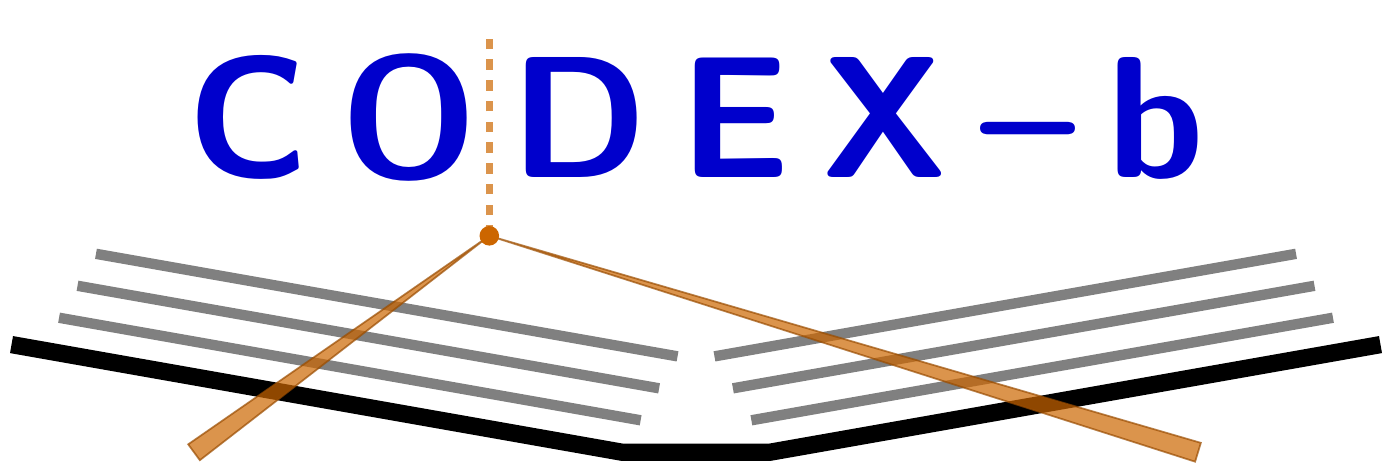}}

\maketitle

{
\fontsize{8}{8}\selectfont 
\columnsep20pt
	\tableofcontents
}

\section{Introduction}
The primary LHC experiments (ATLAS, CMS, LHCb, ALICE) are scheduled for ongoing and upcoming upgrades and data collection until at least 2038.
A central component of the (HL-)LHC program will be searches for dark or hidden sectors Beyond the Standard Model (BSM).
Displaced decays-in-flight of exotic long-lived particles (LLPs) are a compelling signature of such sectors and 
generically arise in any theory containing a hierarchy of scales and/or small parameters. 
Both cases are famously realized in the Standard Model (SM), 
in which many decay widths are suppressed by the $m_W\gg \Lambda_{QCD}$ hierarchy, loop and phase-space suppressions, 
and/or the smallness of one or more CKM matrix elements.  
The $K^0_L$, $\pi^\pm$, neutron and muon are the most spectacular examples of microscopic particles naturally acquiring a very long decay length.
Such LLPs are also ubiquitous in BSM scenarios featuring \emph{e.g.}~Dark Matter, Baryogenesis, Supersymmetry or Neutral Naturalness. 

\begin{figure}
\includegraphics[width=0.45\textwidth]{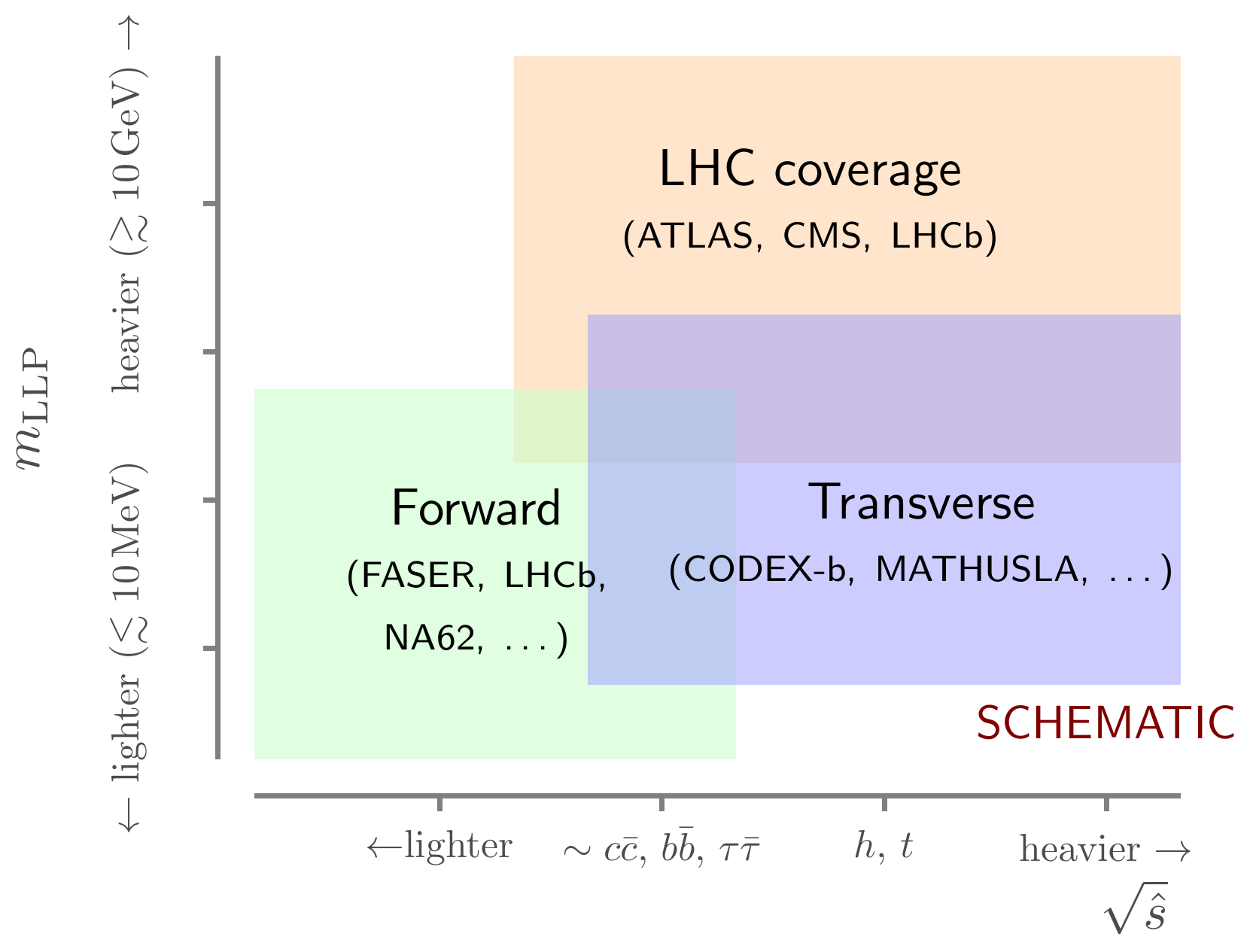} 
\caption{Complementarity of different experiments searching for LLPs~\cite{Aielli:2019ivi}.}
\label{fig:schematic}
\end{figure}

The program to search for LLPs at ATLAS, CMS and LHCb is vibrant and draws on the expertise of both analysis and detector specialists, 
as well as theorists \cite{Alimena:2019zri}. 
The sensitivity of both ATLAS and CMS to the decay-in-flight of LLPs is greatest when they are relatively heavy (\mbox{$m\gtrsim 10$ GeV}),
though there are some important exceptions (e.g.~\cite{CMS:2021sch,CMS:2021juv}).
The reason for this is that backgrounds and trigger challenges can strongly limit the reach for light LLPs in the complicated environment inherent to a high-energy, high-intensity hadron collider. 
These difficulties are offset to a large degree by LHCb and FASER, thanks to, in the former case, its high-precision VErtex LOcator (VELO) and, in the latter case, its large amount of shielding.
Because of their locations and geometry, their sensitivity is restricted to relatively short lifetimes and production at low center-of-mass energies,
and their sensitivity to LLPs produced in, e.g., exotic Higgs or $B$ decays can be quite limited, especially for $c\tau \gtrsim 1\,\mathrm{m}$. 
To achieve comprehensive coverage of the full LLP parametric landscape, one or more high volume, \emph{transverse} LLP detectors are therefore needed (see Fig.~\ref{fig:schematic}). 
\CODEXb~(\emph{``COmpact Detector for EXotics at LHCb"}) is a low cost option, which makes use of existing technology and infrastructure.

The \CODEXb experiment is a special-purpose detector proposed to be installed near the LHCb interaction point to search for displaced decays-in-flight of exotic LLPs~\cite{Gligorov:2017nwh,Aielli:2019ivi,Dey:2019vyo}.
A recent Expression of Interest (EoI) presented the physics case and extensive experimental and simulation studies for the proposal~\cite{Aielli:2019ivi}.
The core advantages of \CODEXb are
\begin{enumerate}[a)]
\item its competitive sensitivity to a wide range of BSM LLP scenarios, exceeding or complementing the sensitivity of existing or proposed detectors;
\item a zero background environment, as well as an accessible experimental location with many of the necessary services already in place;
\item its ability to tag events of interest within the existing LHCb detector, independently from the LHCb physics program;
\item its compact size and consequently modest cost, with the realistic possibility to extend detector capabilities for neutral particles in the final state.
\end{enumerate}
\CODEXb will provide competitive sensitivity over a large range of different LLP production and decay mechanisms; 
extensive studies of this can be found in the expression of interest~\cite{Aielli:2019ivi}
and are outlined in brief below.

\begin{figure}[t!]
  	\includegraphics[width = \linewidth]{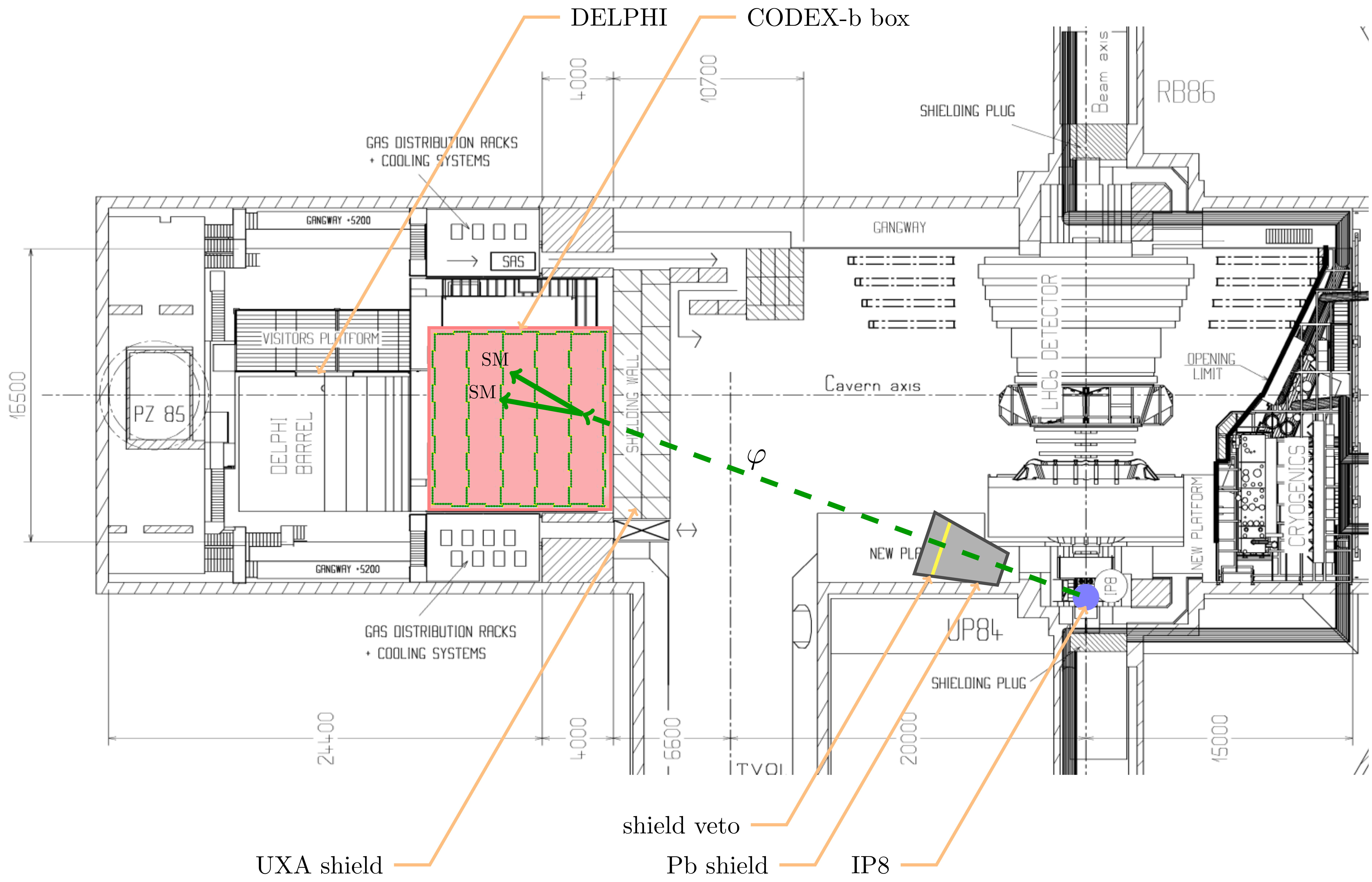}
	\caption{Top view layout of the LHCb experimental cavern UX85 at point 8 of the LHC, overlaid with a top view schematic of the \CODEXb detector. Adapted from Ref.~\cite{Gligorov:2017nwh}.}
	\label{fig:LHCbCav}
\end{figure}

The proposed \CODEXb detector would be located roughly 25 meters from the LHCb interaction point (IP8)
and have a nominal fiducial volume of $10\times10\times 10\,\mathrm{m}^3$ (see Fig.~\ref{fig:LHCbCav}). 
The location roughly corresponds to the pseudorapidity range $0.13<\eta<0.54$. 
Backgrounds are controlled by passive shielding provided by the existing concrete UXA radiation wall, combined with an array
of active vetos and passive shielding to be installed adjacent to IP8.

A smaller proof-of-concept demonstrator detector, \CODEXbeta, will be operated during Run~3 of the LHC, with installation planned for the winter of 2022--2023. 
This detector will be placed in the proposed location of \CODEXb, shielded only by the existing, concrete UXA wall. 

The remainder of this white paper is structured as follows. We review the motivation and physics reach for CODEX-b in Sec.~\ref{sec:theory}. 
The optimization of the detector geometry and the status of the background simulations are discussed in Secs.~\ref{sec:design} and \ref{sec:background}, 
while Sec.~\ref{sec:codexbeta} describes the status of the \CODEXbeta demonstrator detector. We conclude in Sec.~\ref{sec:conclusions}. 

\section{Physics Case \label{sec:theory}}

Discussed in extensive detail in Ref.~\cite{Aielli:2019ivi}, 
the physics case for \CODEXb is principally motivated by the very broad class of models that may be explored through LLP signatures: 
almost any model with either a hierarchy of mass scales, loop suppressions and/or small couplings may feature an LLP in its spectrum.
The Standard Model (SM) is the most famous and obvious example of such a theory, which has all three of these features,
and many extensions of the SM exhibit at least one. 
The broad array of possibilities raises the problem of how to achieve comprehensive coverage of the theory landscape,
something which can only be accomplished with a set of complementary experiments and searches. 
Given the many possible topologies, some amount of ``theory bias'' is inevitable. 
We have considered two complementary approaches, studying: 
\begin{enumerate}
 \item \emph{Minimal models} or ``portals'', which are simple SM extensions with a single new particle, neutral under all SM gauge interactions. 
 Such simplified models have limited predictive power and physical interpretation.
They are, however, good representatives of more complicated models, 
which aim to address one or more outstanding problems of the Standard Model. 
This approach has led to the development of a set of benchmark models during the Physics Beyond Colliders (PBC) effort \cite{Beacham:2019nyx}.
\item \emph{Complete models}, which are more complicated and aim to address one or more of the outstanding puzzles of the Standard Model.
This includes the hierarchy problem, baryogenesis and dark matter.
\end{enumerate}
In the remainder of this section we briefly summarize our findings for these two lines of reasoning.

\subsection{Minimal models}
Underpinning the minimal model approach is the fact that the symmetries of the SM already strongly restrict 
the possible couplings through which a new, neutral state may interact with the SM sector, 
and a simple classification is possible through the spin of the new state. 
One typically considers a scalar ($S$), pseudo-scalar ($a$), a fermion ($N$) or a vector ($A'$), where each allows for a handful of dimension 4 and/or dimension 5 operators:
\begin{subequations}
\begin{align}
	\text{Abelian hidden sector:\quad}& F_{\mu\nu}F'^{\mu\nu},\quad h A'_{\mu}A'^{\mu}\label{eq:vectorportal}\\
	\text{Dark Higgs:\quad}&S^2H^\dagger H, \quad S H^\dagger H\\
	\text{Heavy neutral leptons:\quad}&\tilde{H} \bar{L}N \label{eqn:hnlportal}\\
	\text{Axion-like particles:\quad}&\partial^\mu a\, \bar \psi \gamma_\mu\gamma_5 \psi,\quad a W_{\mu\nu}\tilde W^{\mu\nu},\nonumber\\ 
 & a B_{\mu\nu}\tilde B^{\mu\nu}, \quad a G_{\mu\nu}\tilde G^{\mu\nu}.\label{eqn:alpportal}
\end{align}
\end{subequations}
Here $F'^{\mu\nu}$ represents the field strength operator to the vector field $A'$;
$H$ the SM Higgs doublet;
$h$ the physical, SM Higgs boson;
$L$ the SM lepton doublets;
$\psi$ any SM fermion;
and $B^{\mu\nu}$, $W^{\mu\nu}$ and $G^{\mu\nu}$ the field strengths of the SM hypercharge, $SU(2)$ and strong forces, respectively.
We also allow for scenarios where a different operator is responsible for the production and decay of the LLP, as summarized below. 

\begin{figure*}[htp]
	\includegraphics[width=0.7\textwidth]{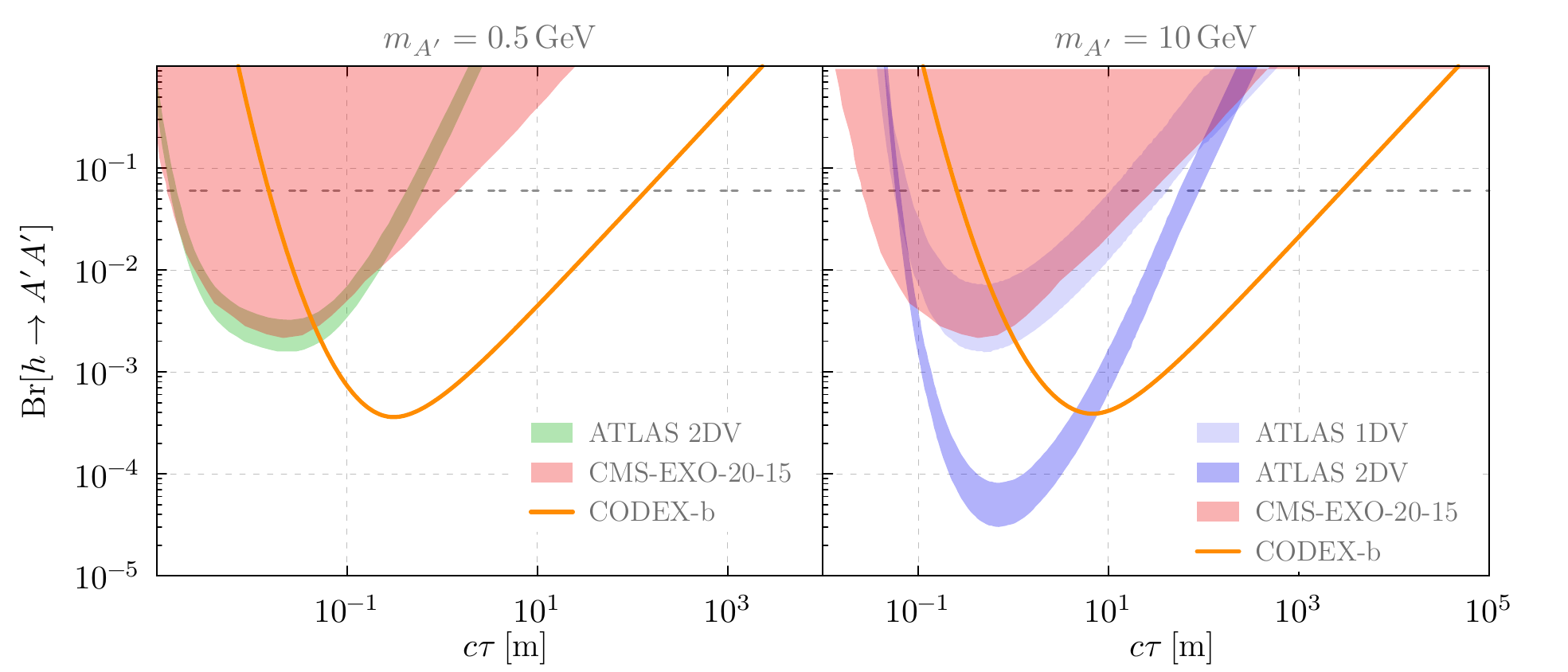}\\
  	\includegraphics[width = 0.42\textwidth]{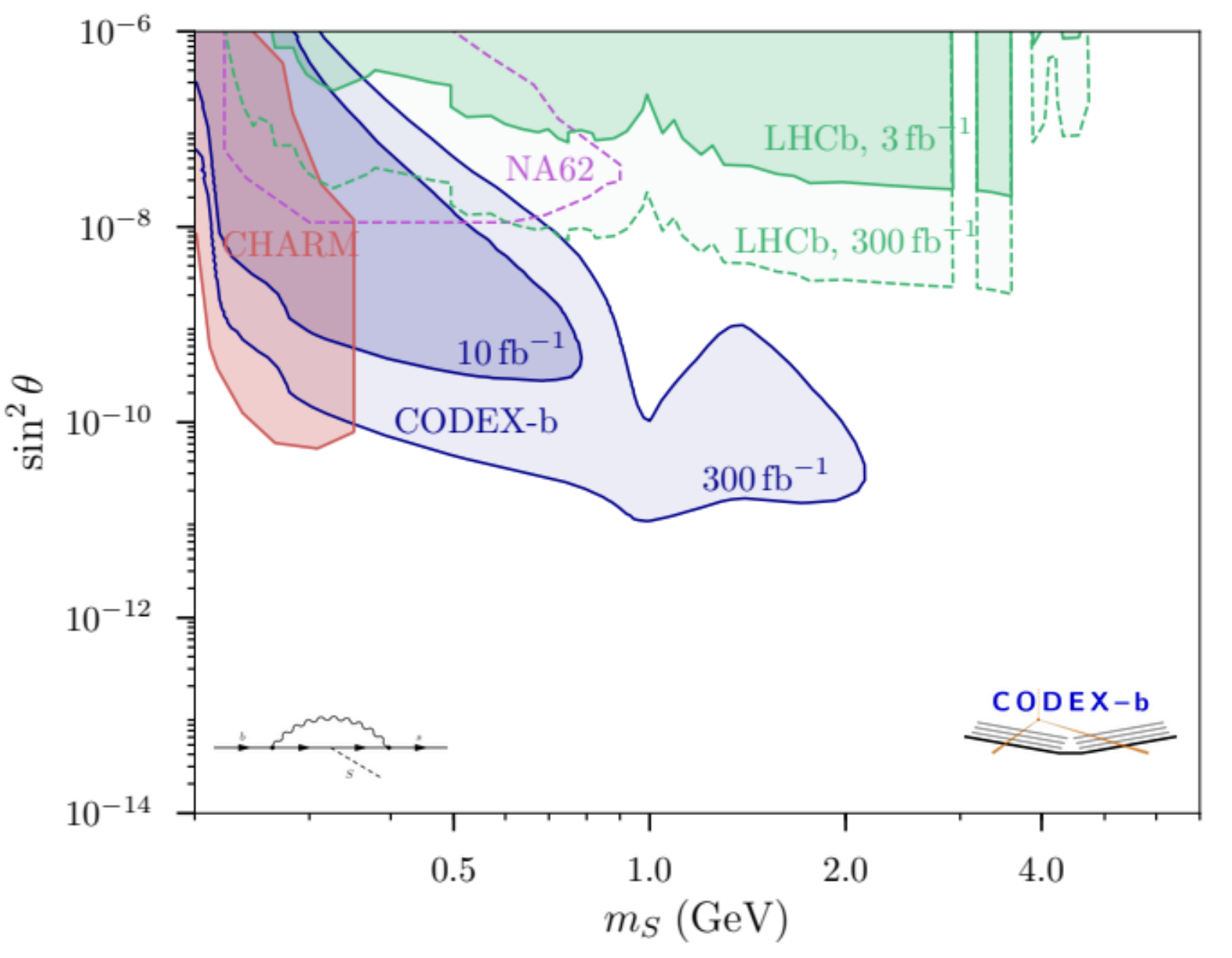}\hspace{15pt}
  	\includegraphics[width = 0.42\textwidth]{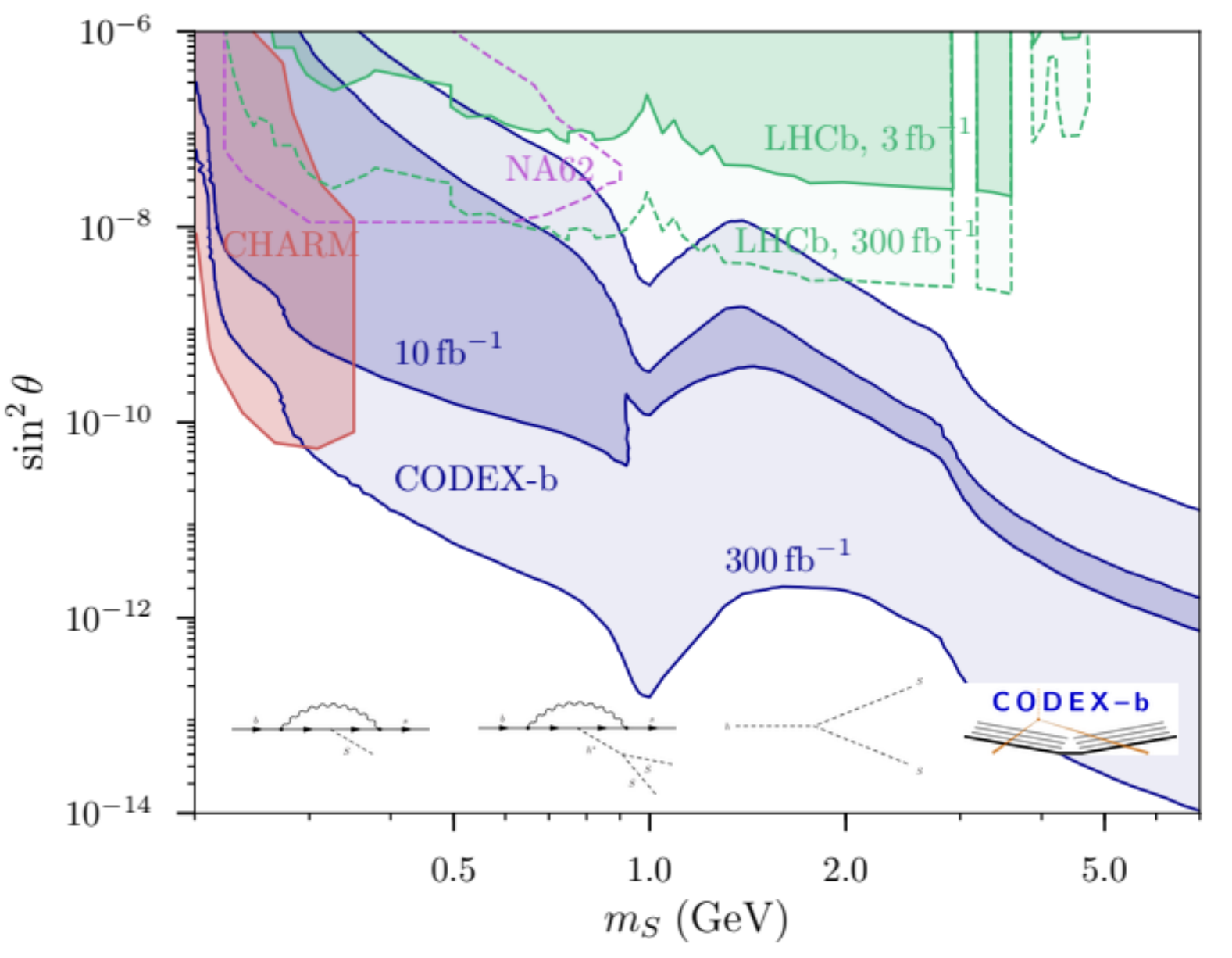}\\
	\hspace{20pt}\includegraphics[width = 0.45\textwidth]{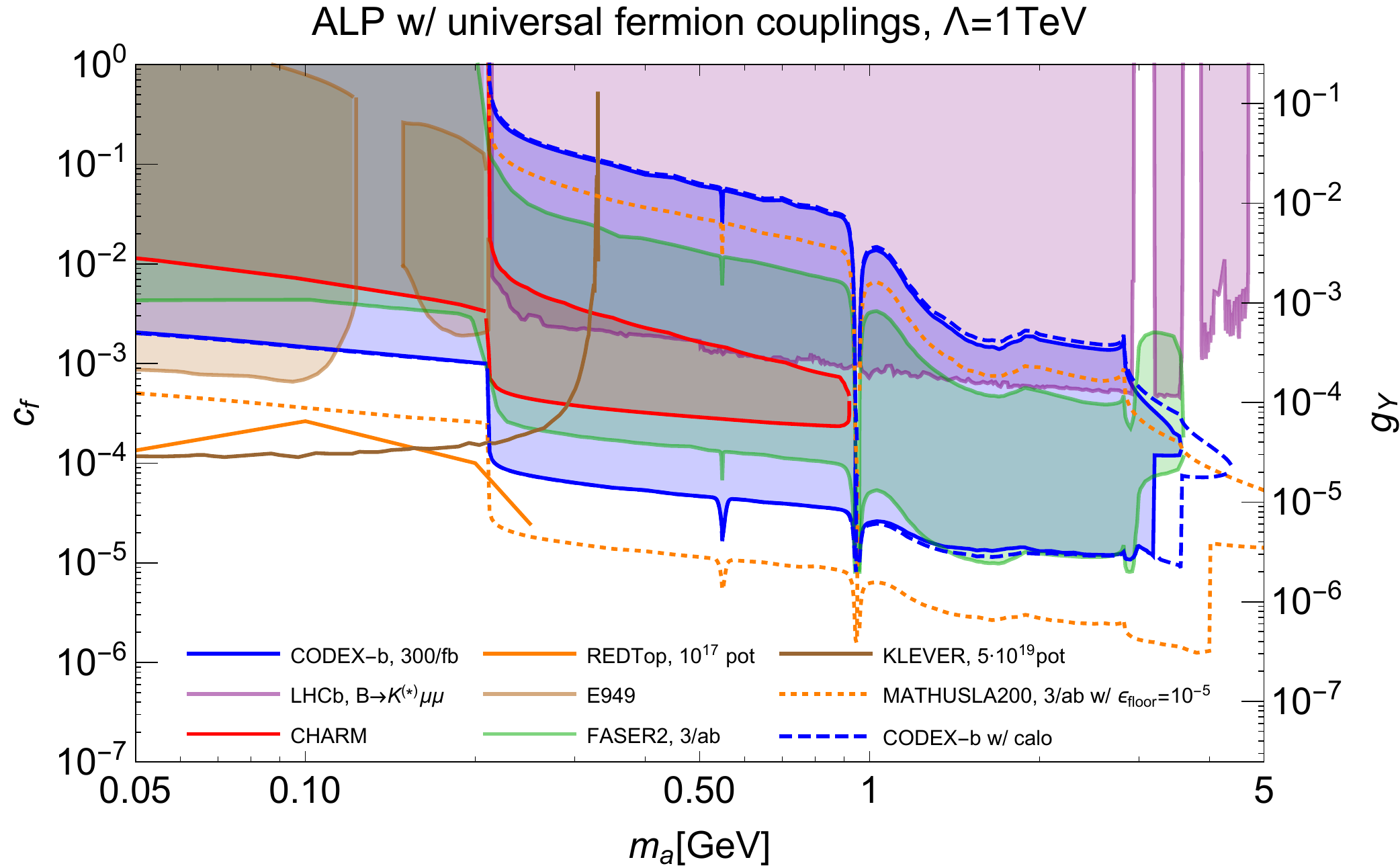}\hspace{10pt}
	\includegraphics[width = 0.45\textwidth]{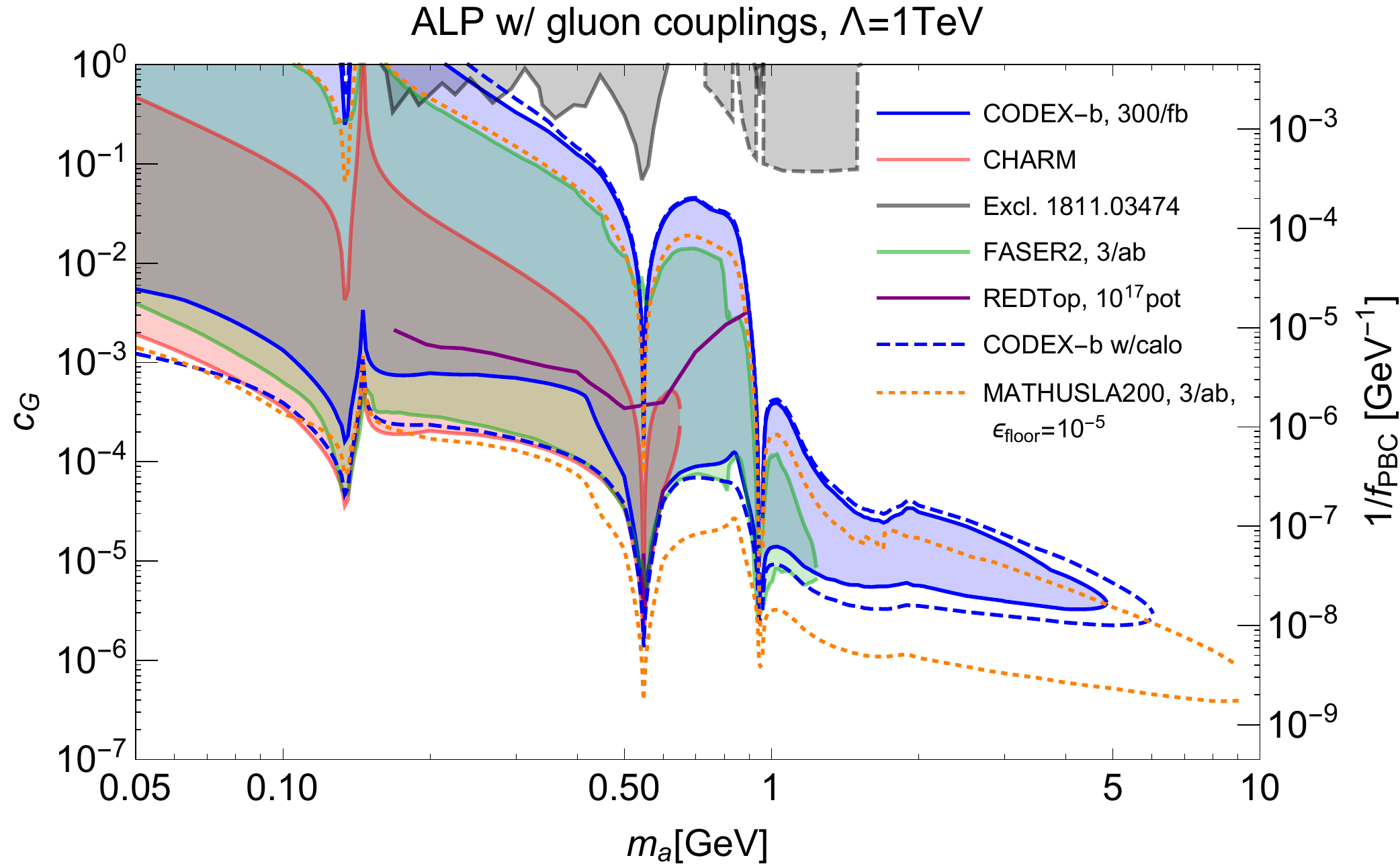}\\
	\includegraphics[width = 0.4\textwidth]{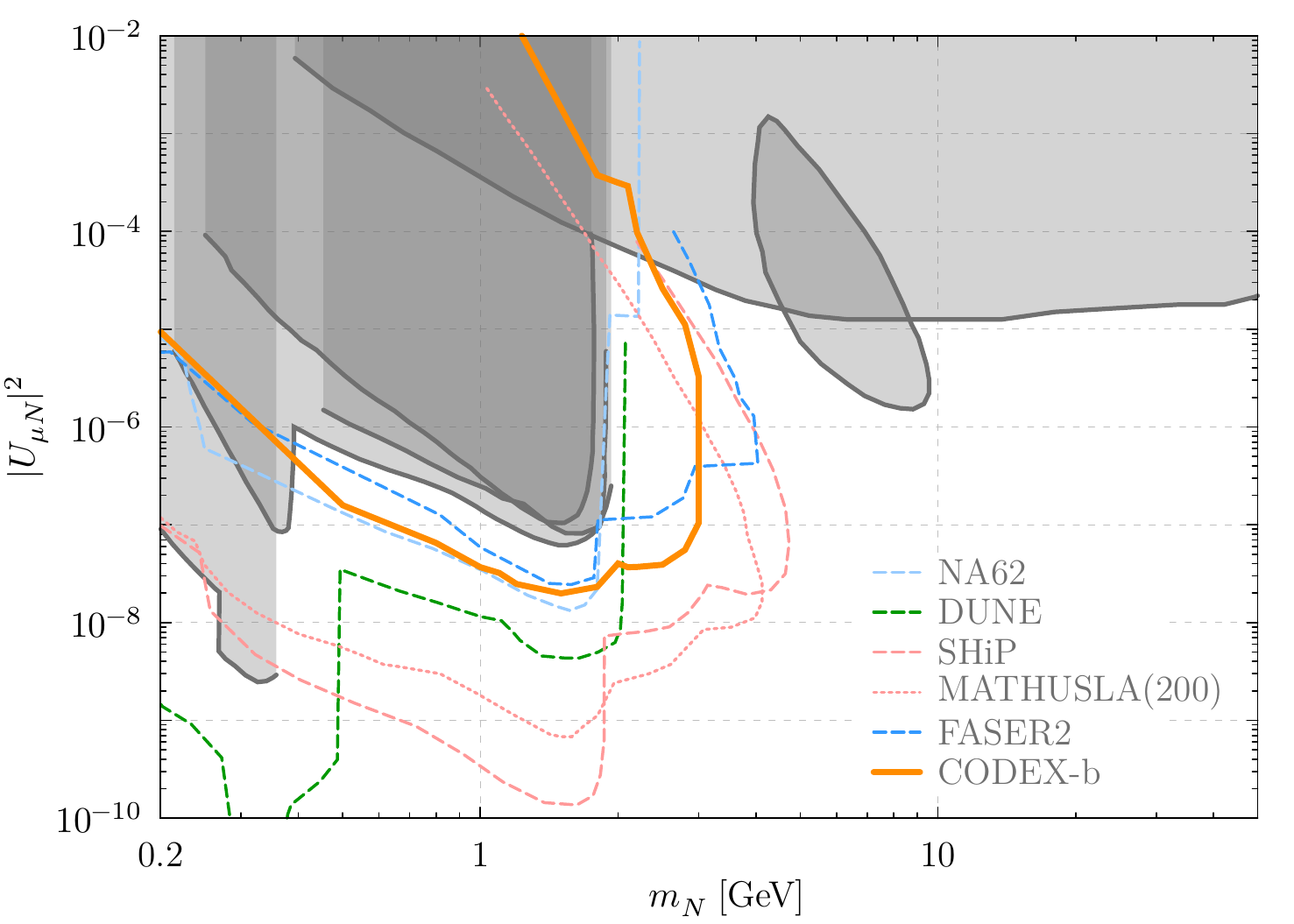}\hspace{20pt}
	\includegraphics[width = 0.4\textwidth]{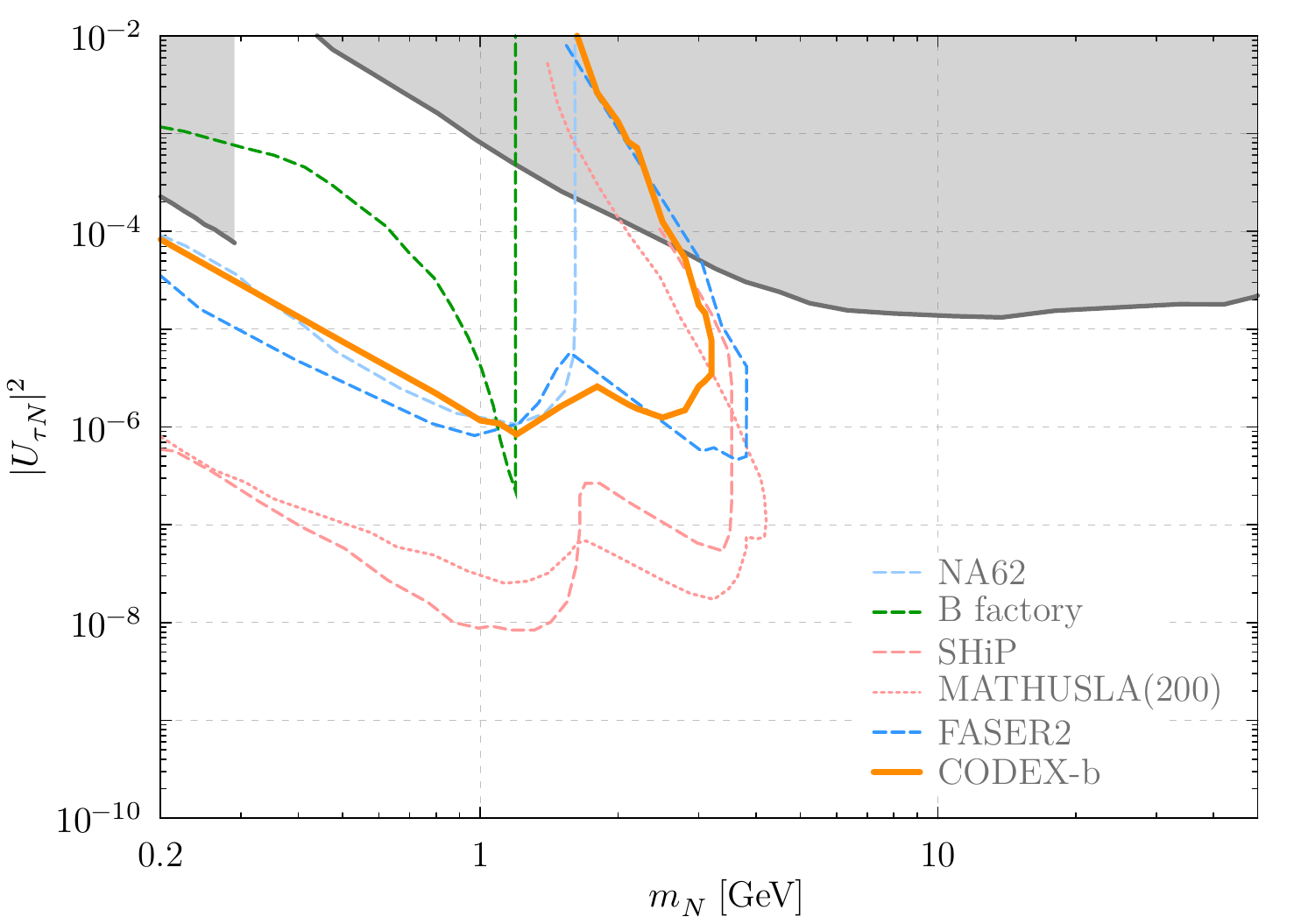}\\
	\caption{
	\textbf{Top:} Reach of \CODEXb for $h\to A'A'$ for two different values of the $A'$ mass, along with the exclusion limit in Ref.~\cite{CMS:2021juv} (red shading); the blue and green shaded bands are expected limits for searches with the ATLAS muon systems, extrapolated to the HL-LHC~\cite{Gligorov:2017nwh}.\quad
	\textbf{Upper Middle:} Projected exclusion power in the dark Higgs simplified model, for the nominal CODEX-b volume with 300 $\text{fb}^{-1}$. 
	The mixed quartic with the SM Higgs was chosen such that $\text{Br}[h\to SS]=0$ ($\text{Br}[h\to SS]=0.01$) in the left (right) panel~\cite{Aielli:2019ivi}.\quad
	\textbf{Lower Middle:} Reach of \CODEXb for fermion-coupled (left) and gluon-coupled (right) ALPs. See~\cite{Aielli:2019ivi} for more details.\quad
	\textbf{Bottom:} Projected sensitivity of \CODEXb to Dirac heavy neutral leptons coupled to $\mu$ (left) and $\tau$ (right) flavored neutrinos, 
	versus current constraints (gray) and other proposed experiments. 
	See~\cite{Aielli:2019ivi} for more details. }
	\label{fig:portals}
\end{figure*}

The \textbf{Abelian hidden sector} model \cite{Schabinger:2005ei,Gopalakrishna:2008dv,Curtin:2014cca} is a very simple extension of the SM with just one additional, massive $U(1)$ gauge boson ($A'$) and its 
corresponding Higgs boson ($H'$).
(See, e.g., Refs.~\cite{Strassler:2006ri,Barbieri:2005ri,Barger:2007im,Morrissey:2009ur,Curtin:2013fra,Bauer:2018onh,Deppisch:2019ldi} for examples of other models with similar phenomenology.)
The $A'$ and the $H'$ mix with, respectively, the SM photon \cite{Holdom:1985ag,Galison:1983pa} and Higgs boson.
If the latter is heavier than the SM Higgs, it decouples from the phenomenology, 
leaving behind the operators in Eq.~\eqref{eq:vectorportal} in the low energy effective theory. 
The $h A'_{\mu}A'^{\mu}$ operator is responsible for the production of the $A'$, through the exotic Higgs decay $h\to A'A'$, 
while the $A'$ decay proceeds through the kinetic mixing operator $F_{\mu\nu}F'^{\mu\nu}$. 
The production and decay rates of the $A'$ are therefore controlled by independent parameters. 
The top row of Fig.~\ref{fig:portals} shows the reach of \CODEXb for two different values of the $A'$ mass.

The most minimal extension of the SM comprises the addition of a single, real scalar degree of freedom ($S$) that couples to the SM Higgs. 
This scenario is often referred to as the \textbf{dark Higgs} or Higgs portal simplified model. 
The model has three free parameters: the mass ($m_S$), the mixing angle with the Higgs ($s_\theta$) and the mixed quartic coupling with the Higgs ($\lambda_D$).
The mixing angle controls the lifetime of $S$ as well as the production rate through exotic $B$ decays, 
as indicated by the penguin diagram in the inset of the upper middle left-hand panel of Fig.~\ref{fig:portals}. 
The mixed quartic coupling controls the rate for pair production of $S$ both in exotic Higgs and $B$ decays, 
as indicated by the diagrams in the inset of the middle right-hand panel of Fig.~\ref{fig:portals}.
LHCb already has sensitivity to this model \cite{Aaij:2016qsm,Aaij:2015tna}, 
but \CODEXb would greatly extend the reach into the small-coupling/long lifetime regime.

\textbf{Axion-like particles} (ALPs) couple to the SM through dimension-5 operators~\eqref{eqn:alpportal}, 
arising in a broad range of BSM models.
They tend to be light when generated via the breaking of approximate Peccei-Quinn-like symmetries, 
and their suppressed couplings make them excellent LLP candidates.
Long-lived ALPs may couple to quark and/or gluons, and may be copiously produced at the LHC through a variety of mechanisms,
including production during  hadronization of quarks and gluons, production from hadron decays via neutral pseudoscalar meson mixing 
and production from flavor-changing neutral-current bottom and strange hadron decays.
In addition, for gluon-coupled ALPs, of particular importance for transverse LLP experiments 
such as \CODEXb is production by emission in the parton shower, which can lead to notable enhancements in sensitivity.
The projected ALP sensitivities of \CODEXb for fermion and gluon-coupled ALPs are shown in the lower middle row of Fig.~\ref{fig:portals}.

\textbf{Heavy neutral leptons} couple to the SM sector via the lepton Yukawa portal, 
mediated by the marginal operator in Eq.~\eqref{eqn:hnlportal},
or may arise from a wide range of simplified NP models involving higher-dimensional operators.
These may include explanations for the neutrino masses or theories of dark matter to name just two (see below).
UV completions of SM--HNL operators imply an active-sterile mixing $\nu_\ell = U_{\ell j} \nu_j + U_{\ell N} N$, 
where $\nu_j$ and $N$ are mass eigenstates, and the active-sterile mixing $U_{\ell N}$ is a matrix element of the extended PMNS neutrino mixing matrix.
The projected HNL sensitivities of \CODEXb in the single flavor mixing regime for the $\mu$ and $\tau$ neutrinos---\textit{i.e.}~mixing with one flavor only---is 
shown in the bottom row of Fig.~\ref{fig:portals}.

\subsection{Complete models}

So far we have considered simplified models that derive their main motivation from their minimality,
but they are only representiatives for more complete extensions of the Standard Model,
which aim to explain open problems such as the hierarchy problem, baryogenesis or the origin of the dark matter. 
Here we review a few examples of complete models featuring LLPs, for which CODEX-b has the potential to make a discovery; 
for more details we refer to~\cite{Aielli:2019ivi}.

In models of \textbf{R-parity violating supersymmetry}, \CODEXb has the potential to discover a light neutralino, produced through exotic $B$, $D$ decays or $Z$ decays~\cite{Dercks:2018eua,Helo:2018qej}. The sensitivity is fairly independent of the flavor structure of the RPV coupling(s), so long as the total branching ratio to at least two charged tracks is not suppressed. In \textbf{relaxion models}~\cite{Graham:2015cka}, the light scalar $S$ in the dark Higgs model from the previous section can play an important role in stabilizing the electroweak scale~\cite{Flacke:2016szy,Choi:2016luu}. \textbf{Neutral Naturalness} models~\cite{Craig:2014aea,Craig:2015pha} build on the Twin Higgs paradigm~\cite{Chacko:2005pe,Chacko:2005un} and aim to alleviate the hierarchy problem by means of an approximately $\mathbb{Z}_2$ symmetric hidden sectors. Some versions of these models~\cite{Craig:2015pha,Craig:2016kue} are examples of hidden valleys~\cite{Strassler:2006im,Han:2007ae} and produce long-lived particles in exotic Higgs decays. 

Though the dark matter itself must be stable or extremely long-lived, in most models additional, unstable particles are needed to achieve the correct relic density. These extra particles are sometimes predicted to have macroscopic lifetimes and could be detected by CODEX-b. For example, \CODEXb would be sensitive to \textbf{inelastic dark matter} models \cite{TuckerSmith:2001hy,Izaguirre:2015zva}, which produce very soft, displaced signatures in collider experiments  \cite{Berlin:2018jbm}. Dark matter models with \textbf{sterile co-annihilation} \cite{DAgnolo:2018wcn} moreover predict a phenomenology comparable to that of the dark Higgs benchmark in the previous section. \textbf{Asymmetric dark matter} models~\cite{Kaplan:2009ag,Kim:2013ivd,Zurek:2013wia} and \textbf{Strongly Interacting Massive Particles (SIMPs)}~\cite{Hochberg:2014dra,Hochberg:2014kqa,Choi:2017zww,Choi:2018iit} explicitly favor the $\sim$ GeV scale. Because the dark matter must be sequestered from the Standard Model, any additional states in the dark sector tend to have macroscopic lifetimes that could be discovered at \CODEXb. In \textbf{Freeze-in models}~\cite{Hall:2009bx}, the dark matter is never in equilibrium with the SM, which enforces very feeble couplings. While these models do not provide a sharp prior on the dark matter mass, they do generically predict macroscopic displacements that could be accessible to \CODEXb.

Some models of baryogenesis rely on out-of-equilibrium decays in the early universe and predict macroscopic decay lengths in collider experiments such as \CODEXb. \textbf{WIMP baryogenesis} is such an example~\cite{Cui:2012jh,Cui:2014twa}. The baryon asymmetry can also be generated through the \textbf{indirect CP-violation} from heavy flavor baryons~\cite{McKeen:2015cuz,Aitken:2017wie}. These models predict exotic $b$-hadron decays to LLPs, which may be detectable with \CODEXb. 

Finally, the \textbf{heavy neutral lepton} model from the previous section can be part of an explanation for the neutrino masses \cite{Mohapatra:1986bd}, the $\nu$MSM \cite{Asaka:2005an,Asaka:2005pn}, dark matter models \cite{Batell:2017cmf}, or models which aim to address the recent $B$ anomalies \cite{Greljo:2018ogz,Asadi:2018wea,Robinson:2018gza}.

\section{Detector design and optimization\label{sec:design}}

\subsection{Baseline design and drivers}

\begin{figure}[t]
	\includegraphics[width = 6.5cm]{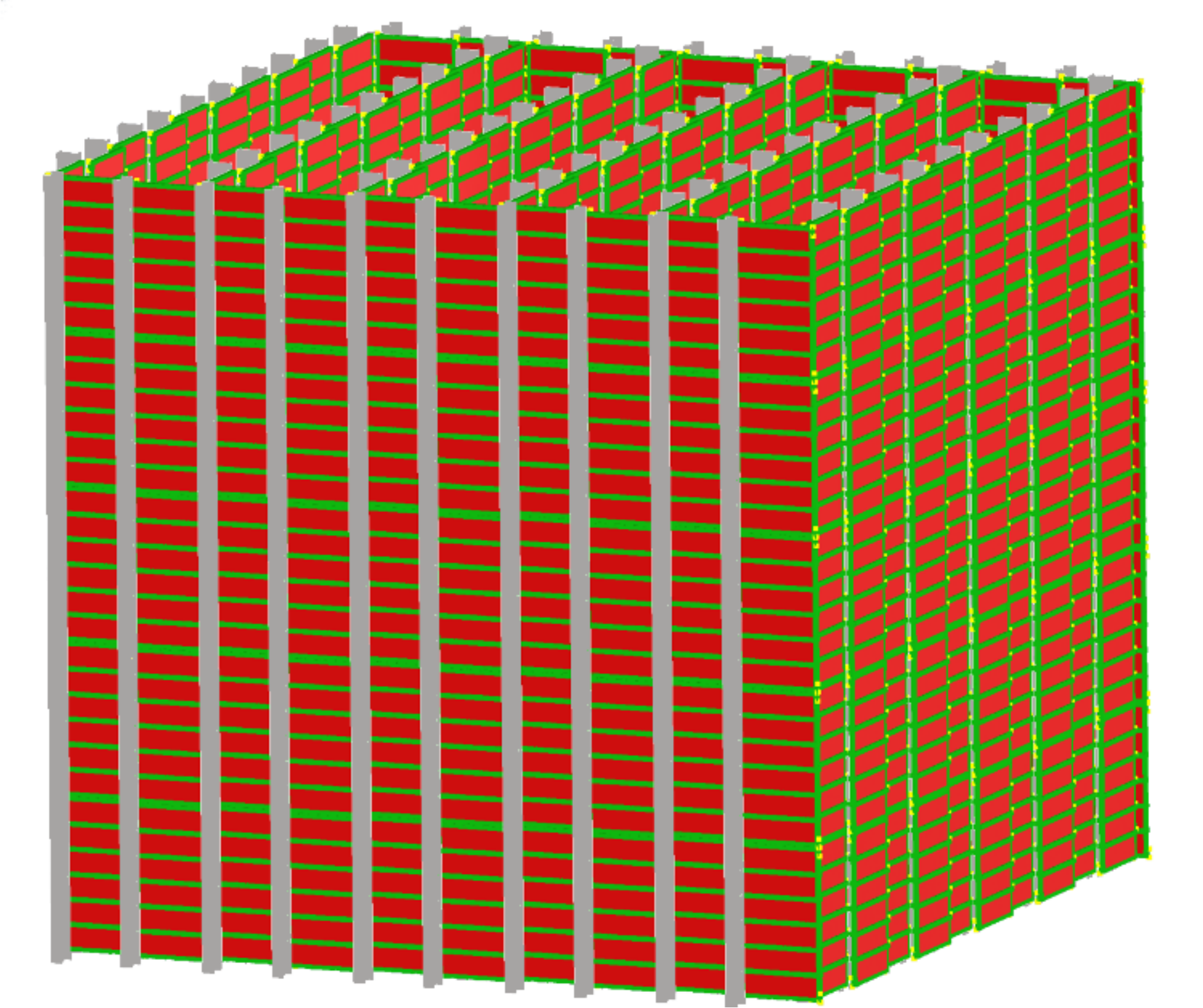}
	\caption{Perspective drawing of the \CODEXb baseline design, containing RPC sextet outer faces and four RPC triplet internal faces. 
	The top faces have been omitted for clarity.
	The RPC triplet panels are shown in red, with green support frames and structural steel supports in dark gray.}
	\label{fig:baseline}
\end{figure}

The baseline configuration considered in Refs~\cite{Gligorov:2017nwh, Aielli:2019ivi} 
comprised sextet RPC panels on the six outer faces of the $10\times10\times10$\,m cubic detector volume,
along with five uniformly-spaced internal stations along the $x$ axis (in beamline coordinates) containing RPC triplet stations.
This cubic volume is located at $x=[26,36]$\,m (transverse), $y=[-7,3]$\,m (vertical) and $z = [5,15]$\,m (forward) with respect to IP8. 
The proposed tracking technology for \CODEXb follows the ATLAS phase-II RPC design~\cite{Collaboration:2285580}, 
so that tracking stations will be composed of arrays of pairs of $1.88\times 1.03$~m$^2$ triplet RPC panels---\textit{i.e.}~the 
fundamental array element is approximately a $2\times2$\,m$^2$ RPC triplet panel---supported 
by a structural steel frame. 
As a result the baseline design has been modified to involve four internal stations, approximately $2$\,m apart.
A perspective technical drawing showing this baseline configuration is shown in Fig.~\ref{fig:baseline}.

The core drivers for an LLP detector design comprise:
\begin{enumerate}
	\item Geometric acceptance: distribution of LLP decay tracks over wide angles motivates the near hermetic coverage of the baseline detector design.
	\item Vertex resolution: Good reconstruction resolution of an LLP decay vertex requires at least six hits per track, 
	with hits as close as feasible to the decay vertex. This motivates the inclusion of internal tracking stations in the baseline design.
	In addition, multiple rescattering of soft tracks means that in practice a minimum track threshold of $600$\,MeV is also imposed,
	and finite hit resolution requires that tracking hits be separated by at least $2$\,cm on any given layer.
	\item Backgrounds: Production of secondary particles in the shielding material (primarily) by muons motivates active veto layers on the front face of the detector.
	Further, vetoing soft cavern backgrounds also requires hermetic coverage of the detector volume.
	In the baseline design these is achieved with sextet RPC stations on the external faces.
\end{enumerate}

While the baseline design has been shown to permit $\mathcal{O}(1)$ track reconstruction 
efficiencies over a broad range of portals~\cite{Gligorov:2017nwh},
it requires a very large amount of instrumented surface:
a total of $400$ $2\times2$\,m$^2$ RPC triplet panels.
The production of the required amount of tracking surface poses a significant challenge, 
both with respect to the short production timescales,  and with respect to costs and installation time.
It is therefore imperative to understand methods that permit minimization (or substantial reduction) of the amount of required tracking surfaces, 
while maximizing (or preserving) the sensitivity to LLP signals. 
For all three design drivers listed above, however, a significant degree of optimization is possible:

First, for instance, while sextet tracking layers are likely obviously required on the entire $x= 36$\,m back face,
far less tracking surface may be required in practice on eg the $z=5$\,m near face, or the $y=3$\,m top or $y=-7$\,m bottom faces.
Because vetoing backgrounds requires good efficiency rather than hit resolution,
much of the external RPC layers might be replaced with cheaper scintillator technologies.

Second, given the above tracking reconstruction requirements, 
internal layers oriented in the $x$-$y$ plane might be more effective at providing coverage of the instrumented volume
than just the four internal faces of the baseline design, 
or might reduce the need for instrumentation on the $y=3$\,m top or $y=-7$\,m bottom faces, 
which pose a more difficult engineering challenge than the vertical faces.

\subsection{Optimization}
\label{sec:opt}
The main challenge in obtaining an optimized detector design is that 
the broad range of BSM scenarios which may be probed leads to a broad range of well-motivated signal morphologies.
As a result, particle reconstruction requirements, efficiencies and acceptances can be quite different from 
portal to portal and model to model, or even across the range of LLP masses and lifetimes.
For instance, the LLP boost distribution and decay products vary significantly 
between the dark Higgs portal benchmark model and the Abelian hidden sector benchmark, mentioned above.

The \CODEXb collaboration has developed a new versatile simulation framework,
that enables fast simulation of the response of variation in the detector geometry and layouts
to different simulated LLP production and decay channels.
With application of optimization techniques,  
preliminary results demonstrate that optimized configurations exist, 
for which \CODEXb can attain good sensitivity over the space of LLP scenarios 
while reducing the required amount of tracking layers by an $\mathcal{O}(1)$ factor.
These results and techniques, to be presented in a forthcoming publication~\cite{Gorordo:2022ip}, 
will permit significant reduction of the forecasted costs, construction and installation times for the experiment.
Moreover, computationally fast estimators of these optimized configurations have been identified.

\begin{figure}[t]
\includegraphics[width=5.5cm]{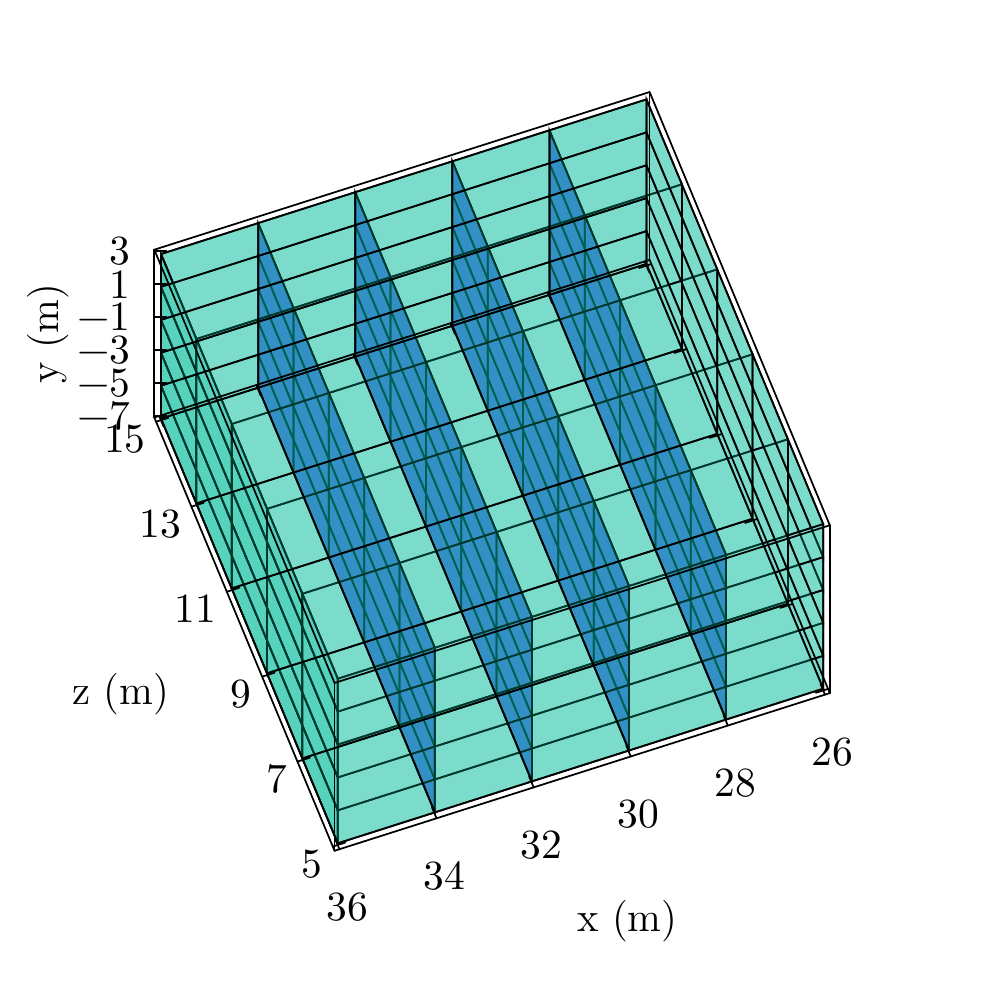}\\
\includegraphics[width=5.5cm]{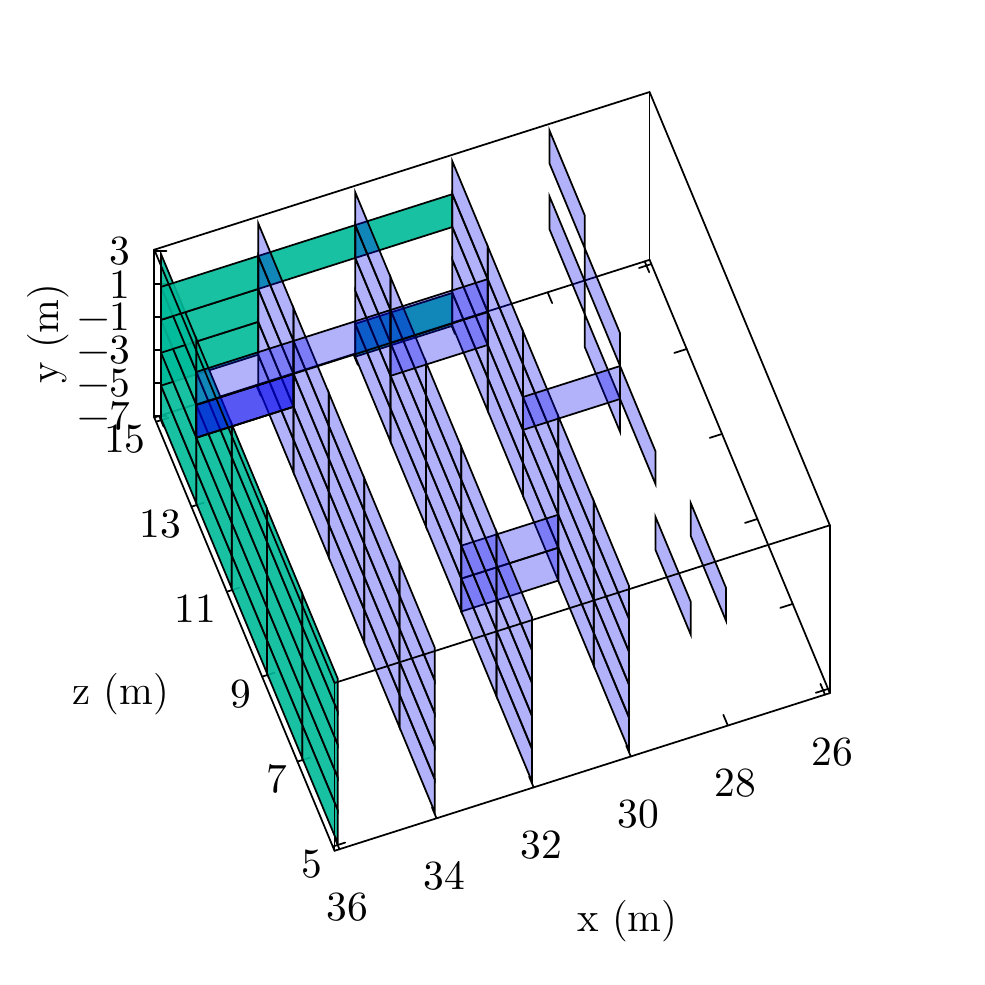}
	\caption{\textbf{Preliminary from Ref.~\cite{Gorordo:2022ip}}. \textbf{Top:} Schematic of the baseline detector geometry, decomposed into $2\times2$\,m$^2$ 
	RPC sextets on the six external faces (green squares) and $2\times2$\,m$^2$ RPC triplets at four internal stations (blue squares),
	for a total of $400$ triplet panels.
	\textbf{Bottom:} An estimator-optimized configuration with only $150$ panels
	(excluding the $x=26$\,m sextet for background rejection), 
	that achieves $\sim50$--$90$\% relative efficiency compared to the baseline, 
	depending on the LLP benchmark.}
	\label{fig:geo}
\end{figure}

\begin{figure}[t]
\includegraphics[width=6.5cm]{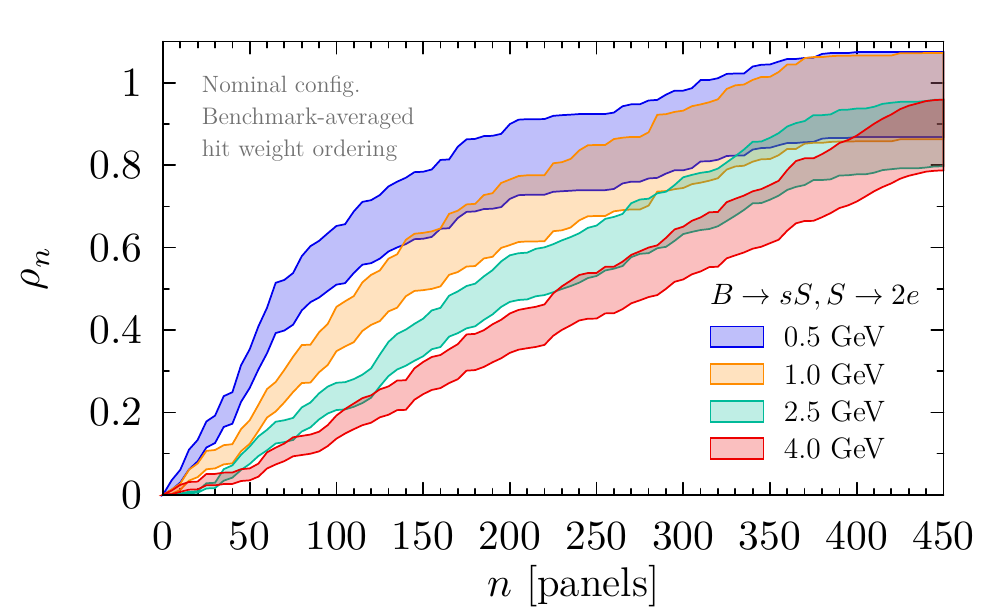}\\
\includegraphics[width=6.5cm]{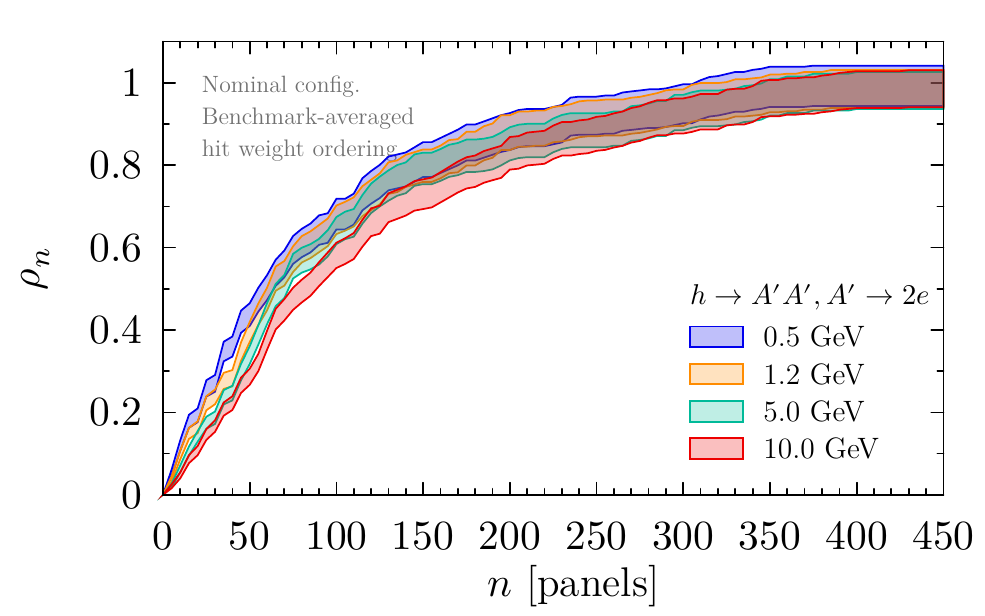}
\caption{\textbf{Preliminary from Ref.~\cite{Gorordo:2022ip}}. 
	Relative vertex reconstruction efficiencies ($1\sigma$ CL bands) as a function of number of panels, 
	as determined by a hit density estimator averaged over an array of dark Higgs and Abelian hidden-sector benchmarks.
	All uncertainties are from MC statistics.}
\label{fig:releffs}
\end{figure}

As a preliminary example, we show in Fig.~\ref{fig:geo} a schematic representation of the baseline configuration,
compared to a reduced configuration containing only $150$ RPC triplet panels---approximately 
$43\%$ of the instrumented surface versus the baseline, excluding the $x=26$\,m face---but 
which can achieve $50$--$90$\% relative vertex reconstruction efficiency, $\rho$, compared to the baseline configuration.
Notably, one sees that the four internal stations along the $x$-axis, and the $x = 36$\,m play a crucial role,
while little sensitivity is gained from many of the external faces.
This configuration is obtained using a simple hit density estimator, 
that tends to be an excellent estimator for systematically-optimized configurations.
More generally, in Fig.~\ref{fig:releffs} we show the relative vertex reconstruction efficiencies as a function of the number of panels,
generated by this estimator, for eleven different benchmarks.
The notable negative curvature for most benchmarks indicates that large reductions in the number of RPC panels are typically possible,
while maintaining good LLP vertex reconstruction efficiencies.
Such results may also be expected to apply for other LLP experiments.

\section{Background and simulations \label{sec:background}}

\subsection{Background Analyses and Shielding Optimization}

\begin{figure}[t]
\centering{
\includegraphics[width = 0.85\linewidth]{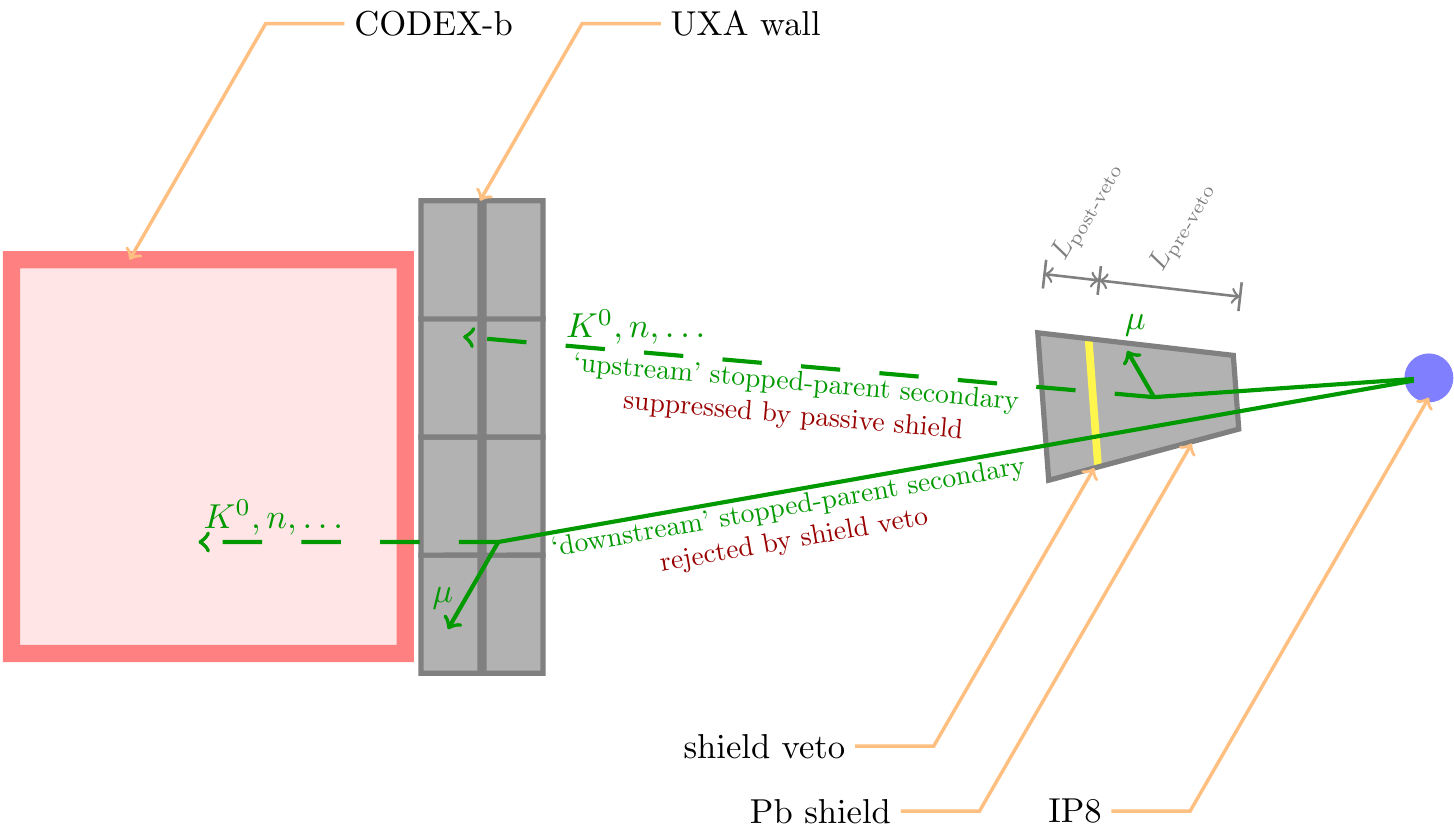}
}
\caption{Cross-section of the shielding configuration of the Pb shield, active shield veto (gold), and concrete UXA wall with respect to IP8 and the detector volume. 
Also shown are typical topologies for production of secondary backgrounds, that are suppressed by shielding or rejected by the veto~\cite{Aielli:2019ivi}.}
\label{fig:shld_cnfg}
\end{figure}

The LHCb interaction point produces a large flux of background primary hadrons and leptons. 
Of these, primary neutral long-lived particles---\textit{e.g.}~(anti)neutrons and $K_L^0$'s---can enter the detector and decay or scatter into tracks resembling a signal decay.
Suppression of these primary hadron fluxes can be achieved with a sufficient amount of passive shielding material:
for a shield of thickness $L$, the background flux suppression $\sim e^{-L/\lambda}$ where $\lambda$ is the material nuclear interaction length.
In the baseline \CODEXb design, the 3\,m of concrete in the UXA radiation wall, corresponding to $7\lambda$ of shielding,
is supplemented with an additional $4.5$\,m of Pb shield, as shown in Fig.~\ref{fig:shld_cnfg}, 
corresponding to an additional $25\lambda$.

However, this large amount of shielding material may act in turn as a source of neutral LLP secondaries, produced mainly by muons or neutrinos that stream through the shielding material and scatter.
The most concerning neutral secondaries are produced $<1$\,m from the back of the shield by muons that slow down and stop before reaching the detector. 
Such muons are therefore invisible to the detector, while their neutral secondaries, such as $K_L^0$'s, may reach the detector volume. 
An example is shown in Fig.~\ref{fig:shld_cnfg}.

\begin{figure}[t]
\centering{
\includegraphics[width = 0.8\linewidth]{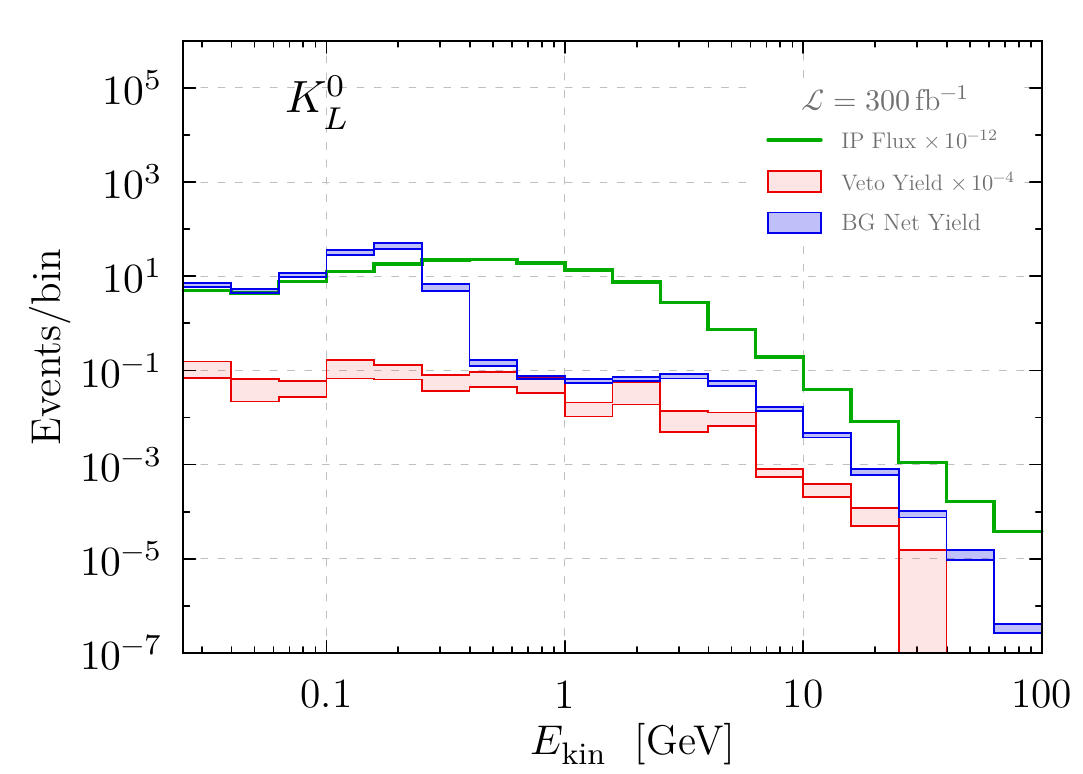}
}
\caption{Example simulated spectrum of secondary $K_L^0$ backgrounds, compared to primary flux (green: $\times 10^{-12}$) 
and the flux vetoed by the active shield (red: $\times 10^{-4}$)~\cite{Aielli:2019ivi}.}
\label{fig:K0L}
\end{figure}

Refs.~\cite{Gligorov:2017nwh, Aielli:2019ivi} have shown that this problem may be solved with the incorporation of an active veto layer 
in the shield itself---the gold layer in Fig.~\ref{fig:shld_cnfg}---placed 
at an optimized location to veto most muons that produce secondaries, but not so close to the IP that all events are vetoed.
Detailed simulations of the setup involve careful treatment of the primary background fluxes at the interaction point, 
folded into a special \texttt{Geant4} simulation of shielding sub elements---usually $\sim 1$\,m thick slices of shielding material---that 
encode the propagation and secondary production of dozens of different background particles species, over a large range of energies. 
An example of the simulated $K_L^0$ flux is shown in Fig.~\ref{fig:K0L}.

The baseline simulation makes a series of conservative assumptions:
\begin{itemize}
    \setlength\itemsep{0mm}
	\item Angular distribution of particle scattering is not exploited;
	all particles scattered within $23^\circ$ of the forward direction are retained.
	\item Detector response to neutral secondary particles is assumed to be 100\%.
	\item Longer path lengths from non-zero angles of incidence on the shield wall are not included.
	\item The active veto is implemented in a single layer, and does not use tracking information.
\end{itemize}
Relaxation of these simulation assumptions would allow for the study of segmented veto layers
that are able to exploit directionality to more efficiently veto background fluxes. 
Further, simulation of the detector response to background fluxes can further improve understanding of likely background-rejection efficiencies.
Both aspects may be studied with the tools already being developed for the optimization studies in Sec.~\ref{sec:opt}, 
with the goal of reducing, possibly substantially, the required amount of lead shielding.

\subsection{Full LHCb-\CODEXb simulation framework}

\begin{figure}
\includegraphics[width = 0.8\linewidth]{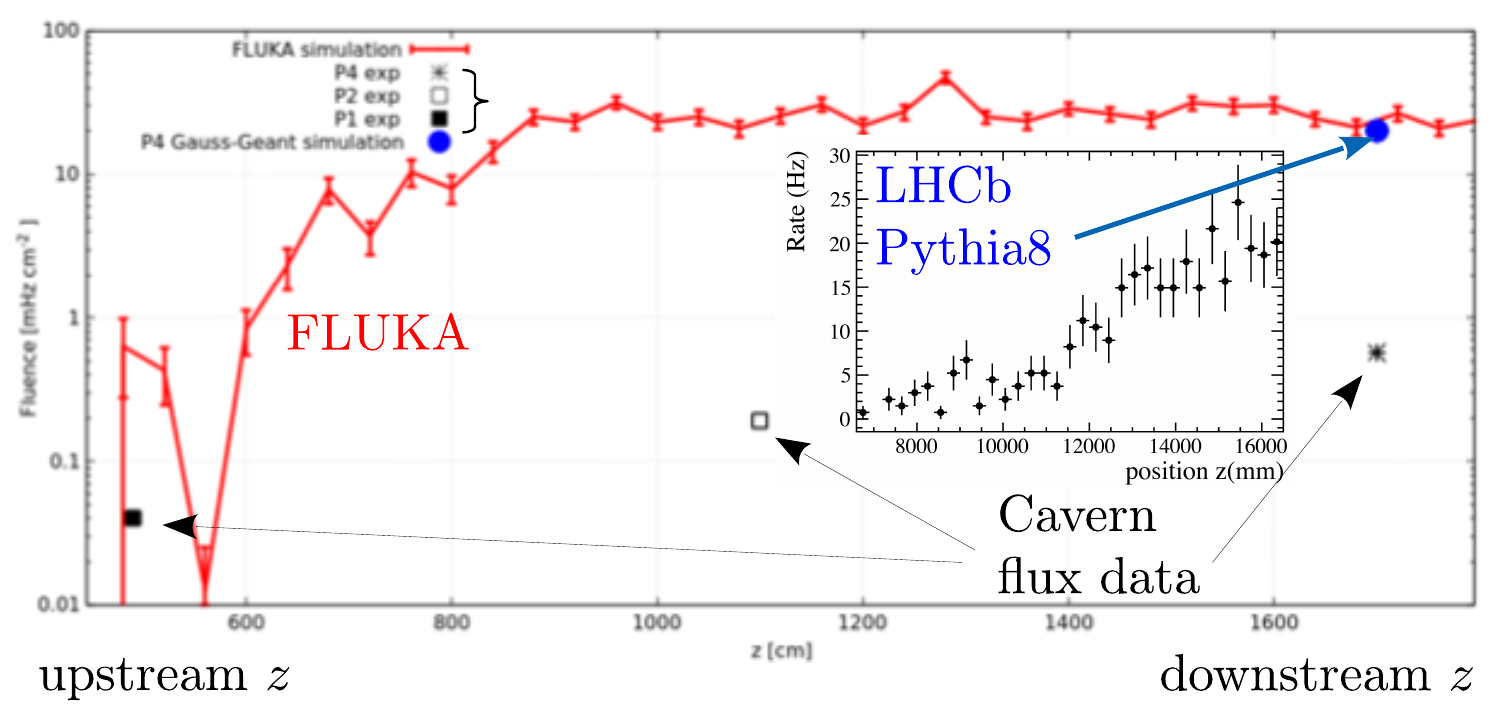}
\caption{Comparison of the charged muon flux behind the concrete shield wall, for Run~2 conditions. 
In red are the \texttt{FLUKA} results from the CERN radiation group, 
while the blue point and the inset are from simulations using the LHCb framework. 
Also marked are the preliminary background measurement data in Ref.~\cite{Dey:2019vyo}.} 
\label{fig:fluka_gauss_compare}
\end{figure}

As mentioned earlier, a salient feature of the \CODEXb proposal is to trigger on events with ``interesting'' pattern of hits in both \CODEXb and the main LHCb detector. 
In addition, details of the cavern infrastructure geometry and the LHCb magnetic field have to be included in the simulation. 
Backgrounds due to processes other than proton-proton collisions at the LHCb interaction point, known as the LHC machine induced background (MIB), also occur. 
To accommodate these, a full simulation, including LHCb, \CODEXb/\CODEXbeta, cavern infrastructure, and MIB is being developed. 
A preliminary setup for Run~1/2 was described in Ref.~\cite{Dey:2019vyo} and is summarized in Fig.~\ref{fig:fluka_gauss_compare}.
This is now being extended to Run~3 data-taking conditions, ATLAS RPCs and the \CODEXbeta geometry (see Sec.~\ref{sec:codexbeta}). 
The work in Ref.~\cite{Dey:2019vyo} retained only the so-called {\tt MCHits} in the sensitive elements from {\tt Geant}: 
This is being extended to include digitization and construction of high-level reconstructed objects (clusters and tracks).   

\subsection{Further background measurements}

The CERN radiation group will be placing a ``BatMon'' (battery operated radiation monitor) unit in the UX85A-D barrack region for Run~3 data taking. 
This will specifically cater to \CODEXb, since all the existing monitors are in the main LHCb cavern and close to the LHCb detector. 
The BatMon unit will complement the charged background flux measurements in Ref.~\cite{Dey:2019vyo}, 
or those that will be measured by \CODEXbeta (see Sec.~\ref{sec:codexbeta}),
since the former is sensitive to thermal neutrons that are difficult to simulate.

\section{CODEX-$\beta$}
\label{sec:codexbeta}

\subsection{Goals}

\begin{table}[tbp]
\newcolumntype{C}{ >{\centering\arraybackslash $} m{3.25cm} <{$}}
\newcolumntype{E}{ >{\centering\arraybackslash $} c <{$}}
\begin{tabular*}{\linewidth}{@{\extracolsep{\fill}}E|CC}
\hline
\text{Tracks} & \text{Total} & K^0_L~\text{contribution}\\
\hline\hline
1	&		(3.87 \pm 0.11) \times 10^{8}	&	(2.94 \pm 0.07) \times 10^{8}	 \\
2	&		(4.09 \pm 0.13) \times 10^{7}	&	(3.74 \pm 0.13) \times 10^{7}	 \\
3	&		(5.96 \pm 1.01) \times 10^{5}	&	(2.92 \pm 0.45) \times 10^{5}	 \\
4+	&		(9.34 \pm 2.10) \times 10^{4}	&	(7.03 \pm 1.91) \times 10^{4}	 \\
\hline
\end{tabular*}
\caption{Total neutral and $K^0_L$ multitrack production during Run~3 in the \CODEXbeta volume for total luminosity $\mathcal{L} = 15\text{fb}^{-1}$, 
requiring $E_{\text{kin}} > 0.4\text{GeV}$ per track~\cite{Aielli:2019ivi}.}
\label{tab:bkg-tracks}
\end{table}

The \CODEXbeta detector is a small-scale demonstrator for the full-scale CODEX-b detector. 
The primary design goal of \CODEXbeta is therefore to validate the key concepts which justify the building and operation of \CODEXb.  Specifically:
\begin{enumerate}
\item  Validate the preliminary background estimates from the \CODEXb  proposal and the 2018 background measurement campaign~\cite{Dey:2019vyo}, 
demonstrating that \CODEXb  can be operated as a zero-background experiment (see Sec.~\ref{sec:background}). 
\item Demonstrate the seamless integration of the detector with the LHCb readout, 
so that candidate LLP events in \CODEXb can be tagged with the corresponding LHCb detector information to aid in their interpretation.   
This is a feature unique to \CODEXb  because LLP detectors linked to ATLAS or CMS must, by necessity, implement hardware triggers.
\item Demonstrate the suitability of Resistive Plate Chambers (RPCs) as a baseline tracking technology for \CODEXb 
in terms of spatial granularity, hermeticity, and timing resolution.
\item  Demonstrate an ability to reconstruct known SM backgrounds expected to decay inside UX-85A 
(the proposed location for \CODEXb  and \CODEXbeta) 
and validate a full simulation of the LHCb detector and cavern environment with these measured backgrounds.
\item  Demonstrate the suitability of the mechanical support required for these RPCs and its scalability to the full \CODEXb  detector.
\end{enumerate}

In particular, observing long-lived SM particles decaying inside the detector acceptance will allow us 
to calibrate the detector reconstruction and the RPC timing resolution.  
The most natural candidates are $K^0_L$ mesons.  
Tab.~\ref{tab:bkg-tracks} summarizes the expected multitrack production from decay or scattering on air-by-neutral fluxes entering \CODEXbeta, 
as well as the contribution from just $K^0_L$ mesons entering the detector. 
Approximately a $\text{few} \times 10^7$ $ K^0_L$ decays to two or more tracks are expected in the \CODEXbeta volume per nominal year of data taking in Run 3, 
so that \CODEXbeta will have the opportunity to reconstruct a variety of $K^0_L$ decays.  
Measurement of the decay vertex and decay product trajectories will allow the boost of the LLP to be reconstructed independently from the time-of-flight information.  
Moreover, measurement of the distribution of $K^0_L$ decay vertices can be compared to expectations from the background simulation 
of the expected $K^0_L$ boost distribution folded against the $K^0_L$ lifetime, 
allowing calibration and validation of our detector simulation and reconstruction. 
Conversely, one may combine the predicted boost distribution and measured vertex distribution to extract the $K^0_L$ lifetime itself.

\begin{figure}[h]
    \begin{center}
        \includegraphics[width=0.45\textwidth]{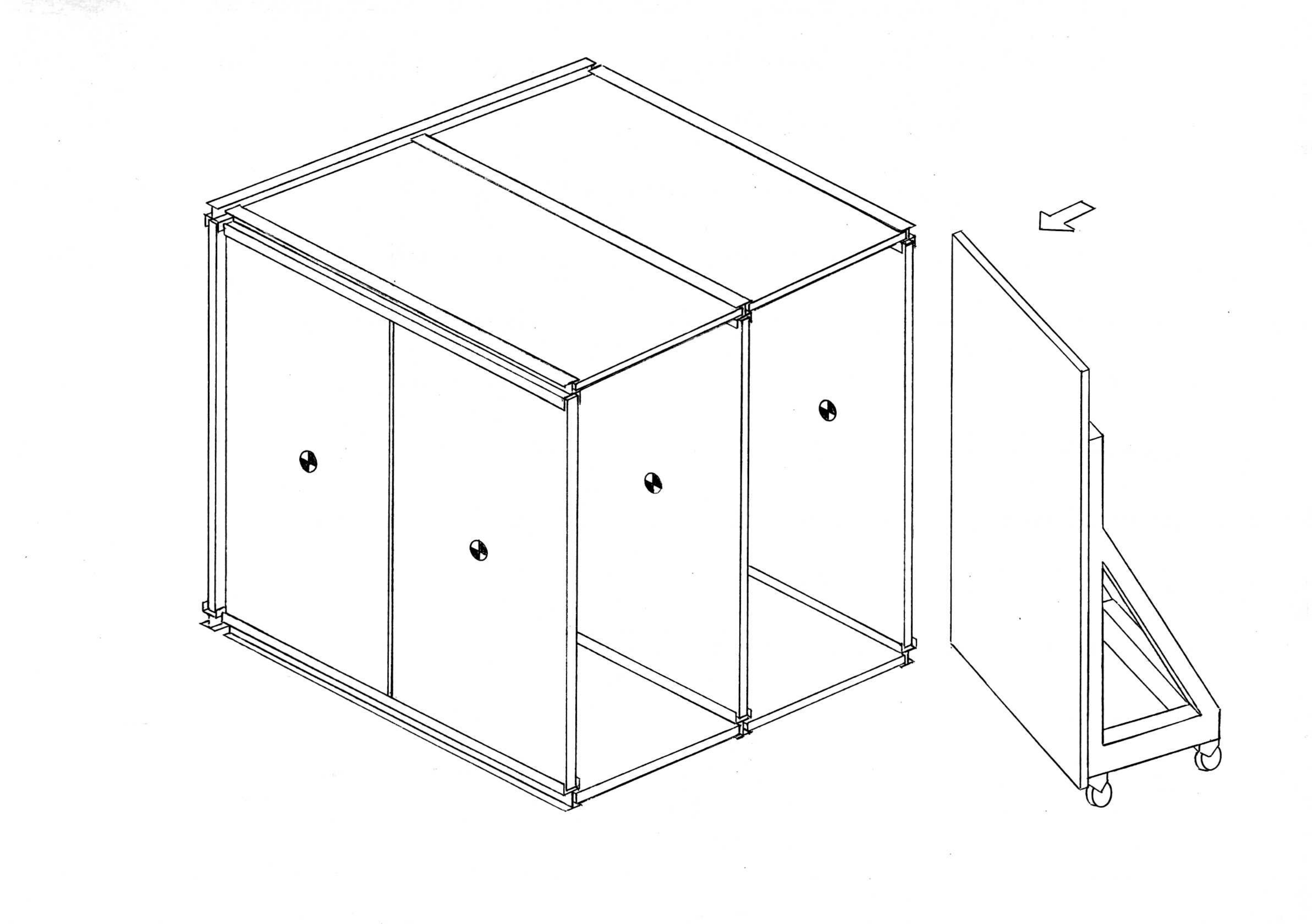}\\
        \includegraphics[width=0.45\textwidth]{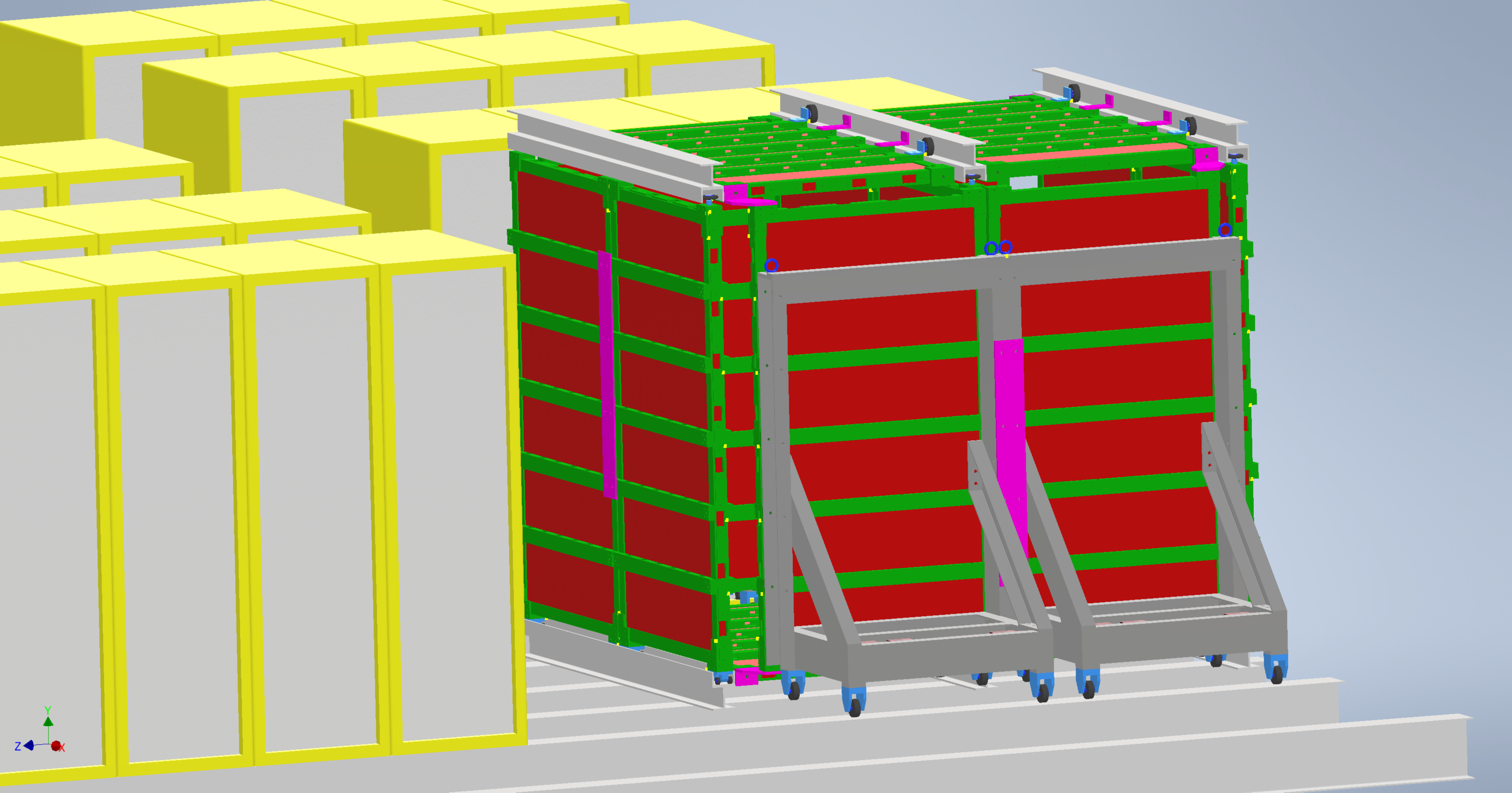}
    \end{center}
    \caption{
        \textbf{Top:} Sketch of \CODEXbeta showing the mechanical framework for the RPC modules; the arrow indicates the direction of incoming particles;the lower left module which will also be on a rolling cart, 
        is not shown for clarity. \textbf{Bottom:} Diagram of \CODEXbeta (green, red, gray) located between the server racks (yellow, gray).}
    \label{fig:demosketch}
\end{figure}

\subsection{Design}

\begin{figure}[tbp]
    \centering
        \includegraphics[width=0.45\textwidth]{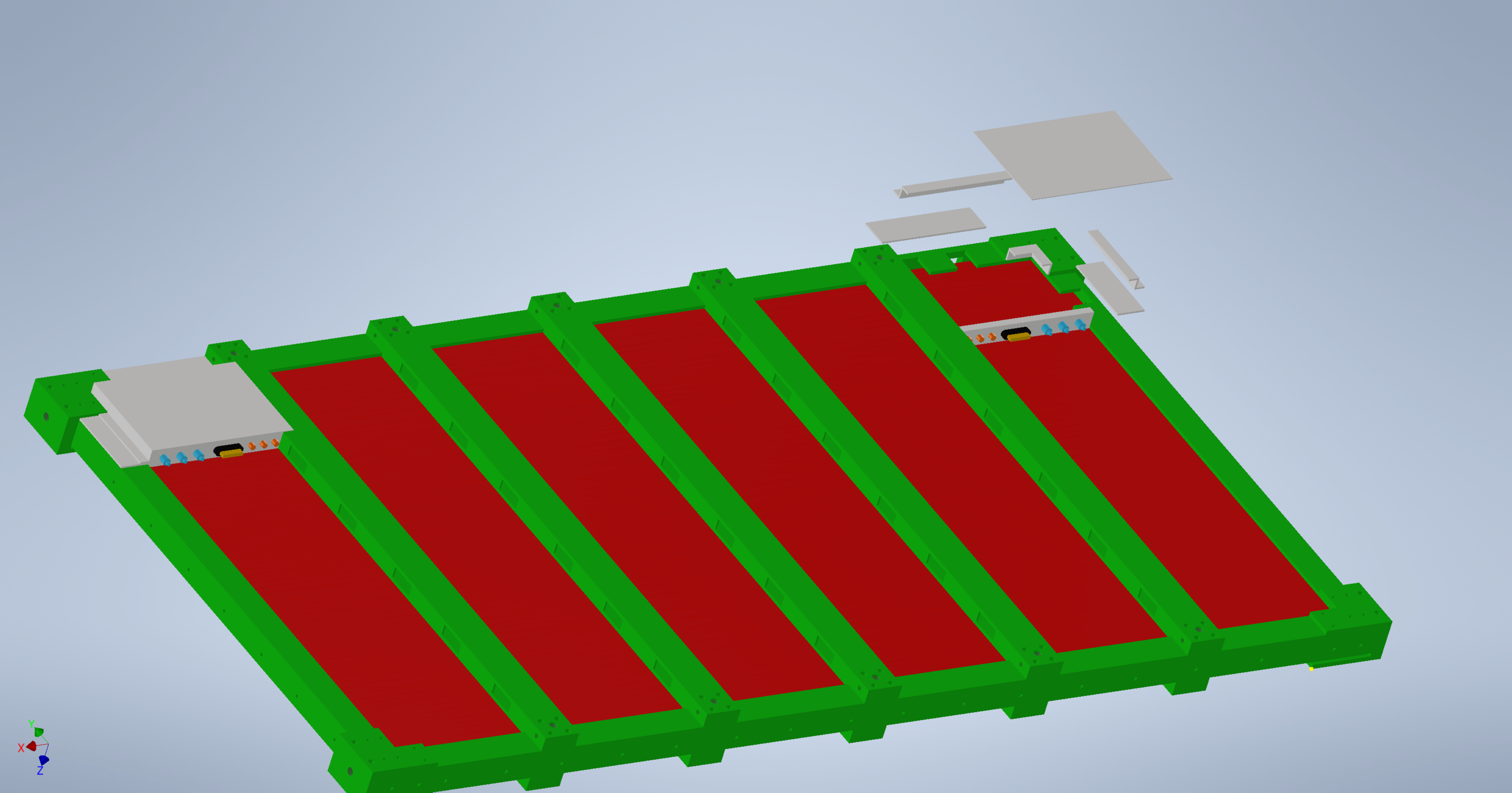}\\
    \includegraphics[width=0.4\textwidth]{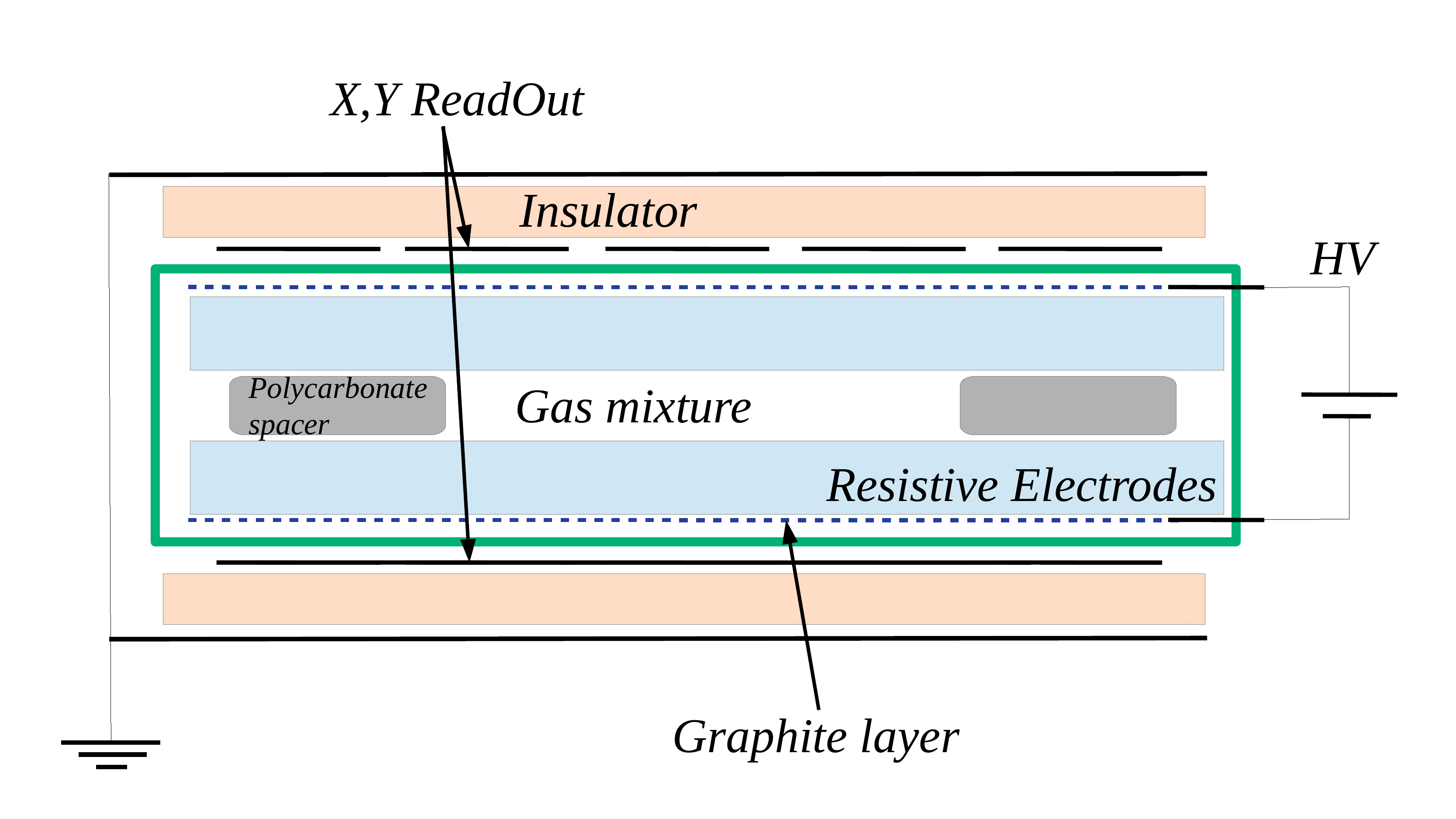} 
    \caption{\textbf{Top:} Diagram of a \CODEXbeta module, including the RPC triplet (red) support frame (green) and service boxes (gray). \textbf{Bottom:} Schematic of the structure of a singlet RPC.}
    \label{fig:rpc}
\end{figure}

As shown in the top of Fig.~\ref{fig:demosketch}, \CODEXbeta will comprise a $2\times2\times2$\,m$^3$ cube with an additional face spanning the interior.   
Each face will contain two modules, 
which will each house a stacked triplet of $2\times1\, \text{m}^2$ RPCs integrated into a self-contained mechanical frame, named CX1,
requiring a total of $(6 + 1)\times2\times3 = 42$ RPC singlets integrated into a total of $14$ modules. The mechanical frame is shown in the top diagram of Fig.~\ref{fig:rpc}, and is specifically designed to withstand the stresses of handling and mounting.

Installation of \CODEXbeta will be challenging as the detector location is within a very confined space, shown in the bottom diagram of Fig.~\ref{fig:demosketch}. Consequently, the \CODEXbeta frame has been designed to be highly modular, and can be assembled with only fastening hardware and no welding required. One of the key installation features is support carts which will be used to move the modules to the frame. The final two modules will remain in the rolling carts and allow internal access to the detector.

The RPCs are read out on both sides by two panels of orthogonal strips, 
with strip pitches of $20$--$25\text{mm}$,
providing pseudorapidity ($\eta$) and azimuth ($\phi$) coordinates. The two service boxes for these readouts are illustrated by the gray boxes in the top diagram of Fig.~\ref{fig:rpc}.
The detector's Faraday-cage design, 
suitable for low-threshold operation,
has been developed to allow a better shielding of the more sensitive front-end electronics.

The chamber is designed as a modular structure, 
where the base element is the singlet RPC, 
composed of the gas gap sandwiched between the $\eta$ and $\phi$ readout strip panels containing the front-end electronics.
Figure~\ref{fig:rpc} shows a schematic of the singlet RPC design.
The fully assembled chamber is composed of three independent singlets, 
comprising a triplet, which are able to provide a three-points track and to work in a self-trigger mode, 
avoiding any external reference system for muon detection.

The resistive electrodes consist of two sheets of phenolic high pressure laminate (HPL), 
with resistivity $\rho$ between $10^{10}$ and $10^{11}\Omega\,\cm$. 
The electrode thickness is $\approx 1.2\,\mm$. 
A matrix of polycarbonate spacers, of $\approx 10\,\mm$ diameter each, 
guarantees the uniformity of the gas-gap thickness for the entire detector area. 
The spacer matrix causes an intrinsic inefficiency of the detector which must be taken into account. 
The spacer lattice has a step of $7\,\cm$, causing a geometric inefficiency of $\approx 1\%$.

The external surface of the electrodes is covered with a paint of graphite with a superficial resistivity $\rho$ of $\approx500k \Omega/\square$, 
which allows a uniform distribution of the high voltage and at the same time allows the inductive pick-up of the charge created within the gas.
The inner surfaces of the electrodes are covered with a very thin layer of linseed oil to avoid the spike effect which creates electrical field 
inhomogeneity and the increasing of the detector noise. 
This is a fundamental component for the detector to work properly.

\begin{figure}[tbp]
    \begin{center}
    \includegraphics[width=0.4\textwidth]{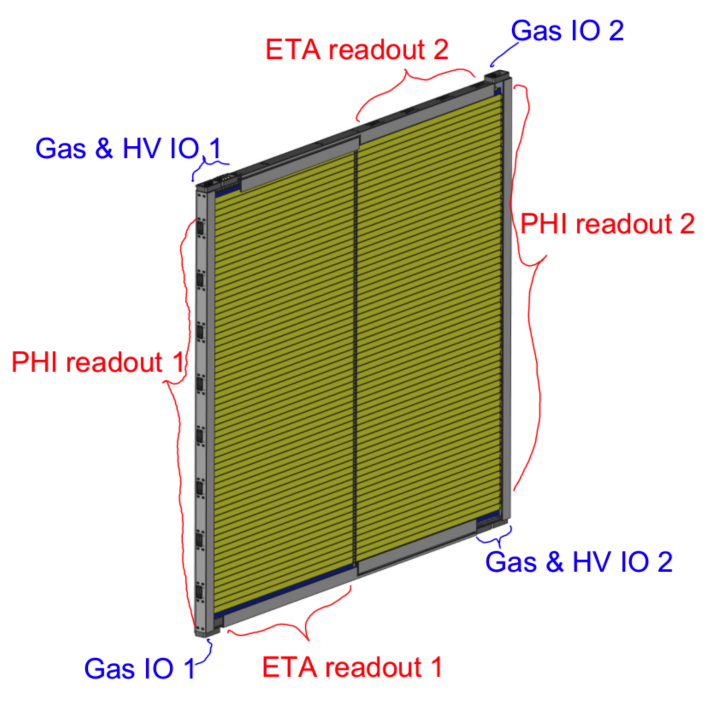}
    \end{center}
    \caption{
        Schematic of a face of \CODEXbeta, comprising two RPC chambers and their support frames.
        (``PHI'' $\equiv \phi$ and ``ETA'' $\equiv \eta$.)
        Each of the two modules includes a support frame.
    }
    \label{fig:RPCTripletSupportBiplab}
\end{figure}

Each module shown in Fig.~\ref{fig:demosketch} comprises a support structure around an RPC chamber and the chamber comprises a stacked triplet of RPC singlets, 
each with an area of $2\times 1\text{m}^2$.
The long and short sides of the chamber are referred to as the $\phi$ and $\eta$ sides, 
respectively, as in Fig.~\ref{fig:RPCTripletSupportBiplab}.  Note that here, the full mechanical frame for each module is not shown.
One $\phi$ side and one $\eta$ side of each module contain readouts for the perpendicular readout strips of the RPCs.
In order to form a full $2\times 2\text{m}^2$ face of the \CODEXbeta cube, 
two modules are placed side-by-side along their $\phi$ sides
such that the readouts are on the opposite outer-edges of the module, 
as in Fig.~\ref{fig:RPCTripletSupportBiplab}. This maintains the hermiticity of the detector. For the full \CODEXb design, the readouts may be moved to the center of the module face to allow for sequential positioning of more than two modules.

\subsection{Timeline}

An initial, detailed timeline, subject to significant change and requiring approval by relevant parties, is given in Fig.~\ref{fig:timeline} in Appendix~\ref{appendix:timeline}, where each block corresponds to a specific task.
The initial primary focus is the development of the CX1 frame to ensure that the second iteration CX2 module frames are built prior to assembly of the RPCs.
Additionally, some of the RPC components have long lead times for procurement.
Already the HPL has been procured, as well as part of the FE boards.
The FOREX procurement depends upon LHCb technical board approval and can be pushed back 
from its current schedule if another dielectric needs to be used.
Actual installation of the detector is expected to begin in early January of $2023$, 
but this will change depending upon the LHCb schedule.
Note that this timeline only covers the hardware aspects of the project and does not include simultaneous work software and reconstructions, as well as \CODEXb development.

\section{Summary and the road ahead\label{sec:conclusions}}

At the time of writing many experimental collaborations are attempting to address the need for experiments dedicated to the study of long-lived particles which can exploit the production cross section and luminosity which the LHC will deliver over the next two decades. The \CODEXb experiment fills a particular role in this ecosystem: relatively compact and affordable, it can be delivered using proven technology in a convenient location which minimises required person power, is able to leverage the existing LHCb data acquisition, and offers competitive or complementary sensitivities for both transverse and longitudinal production of LLPs. The \CODEXb collaboration has validated most of the key assertions made in the 2017 proposal and is today ready to build and commission the \CODEXbeta demonstrator, pending necessary approvals, opening the road towards a construction of the full \CODEXb detector and its deployment in the late 2020s.

Our immediate priority is to build and deploy the \CODEXbeta demonstrator in time to record the majority of Run~3 luminosity available at IP8. This data will allow us to measure both the overall background levels in UX-85A and their spatial distribution with high precision, enabling the active veto shield required by \CODEXb to be optimized in both size and shape. The reconstruction of SM backgrounds will provide invaluable validation of the \CODEXb simulation and allow us to confidently optimize the ultimate spatial and temporal granularity required for the \CODEXb RPCs, as well as their geometric coverage.

By integrating the \CODEXbeta readout into the LHCb data stream, we will validate the data acquisition model and its scalability to the full detector.
Indeed the integration of \CODEXb and LHCb data streams will permit the tagging of events of interest in the LHCb detector. 
If LLP events are detected, this would aid in the determination of the underlying identity of the LLP and its production processes.
Further studies are required to understand the physics potential of this capability for a number of well-motivated scenarios, e.g.~Higgs VBF production and exotic $B$ meson decays, and data taken by \CODEXbeta will be invaluable input to this. 

By the end of 2025 we expect to have a fully optimized geometry for both \CODEXb and its active shield veto, a full specification of required performance for the \CODEXb RPCs, and a concrete design for the final experiment's mechanics. Based on these, we will produce a Technical Design Report and aim for a staged construction and installation process, taking full advantage of the fact that UX-85A is a shielded environment where we can work both during and outside long shutdowns of the LHC. 

\acknowledgements
This work was supported in part by the Laboratory Directed Research and Development Program of Lawrence Berkeley National Laboratory under U.S.~Department of Energy Contract No.~DE-AC02-05CH11231.  VVG acknowledges support of the European Research Council under Consolidator grant RECEPT 724777.

\appendix
\section{Collaboration leadership}
The \CODEXb collaboration has members who are affiliated with ATLAS, CMS, LHCb and the theory community. 
The current leadership of the collaboration:
\begin{table}[!htp]
\newcolumntype{C}{ >{\raggedright\arraybackslash } m{5.5cm} <{}}
\scalebox{0.8}{\parbox{1.2\linewidth}{ 
 \begin{tabular*}{\linewidth}{CC}
Role&Name  \\\hline\hline
Spokesperson & Phil Ilten (U.~Cincinnati)\\
Deputy spokesperson & Vladimir Gligorov (LPNHE)\\
Physics coordination& Xabier Cid Vidal (IGFAE) \\
Deputy physics coordination& Carlos V\'azquez Sierra (CERN)\\
LHCb integration \& commissioning & Daniel Johnson (MIT)\\
Simulation& Biplab Dey (ELTE)\\
Reconstruction& Louis Henry (CERN)\\
Installation \& commissioning& Michael Wilkinson (U.~Cincinnati)\\
Future design& Dean Robinson (LBNL)
\end{tabular*}
}}
\end{table}

\section{\CODEXbeta timeline\label{appendix:timeline}}
\begin{figure*}[h!]
    \includegraphics[width=0.9\textwidth,angle=0,origin=c]{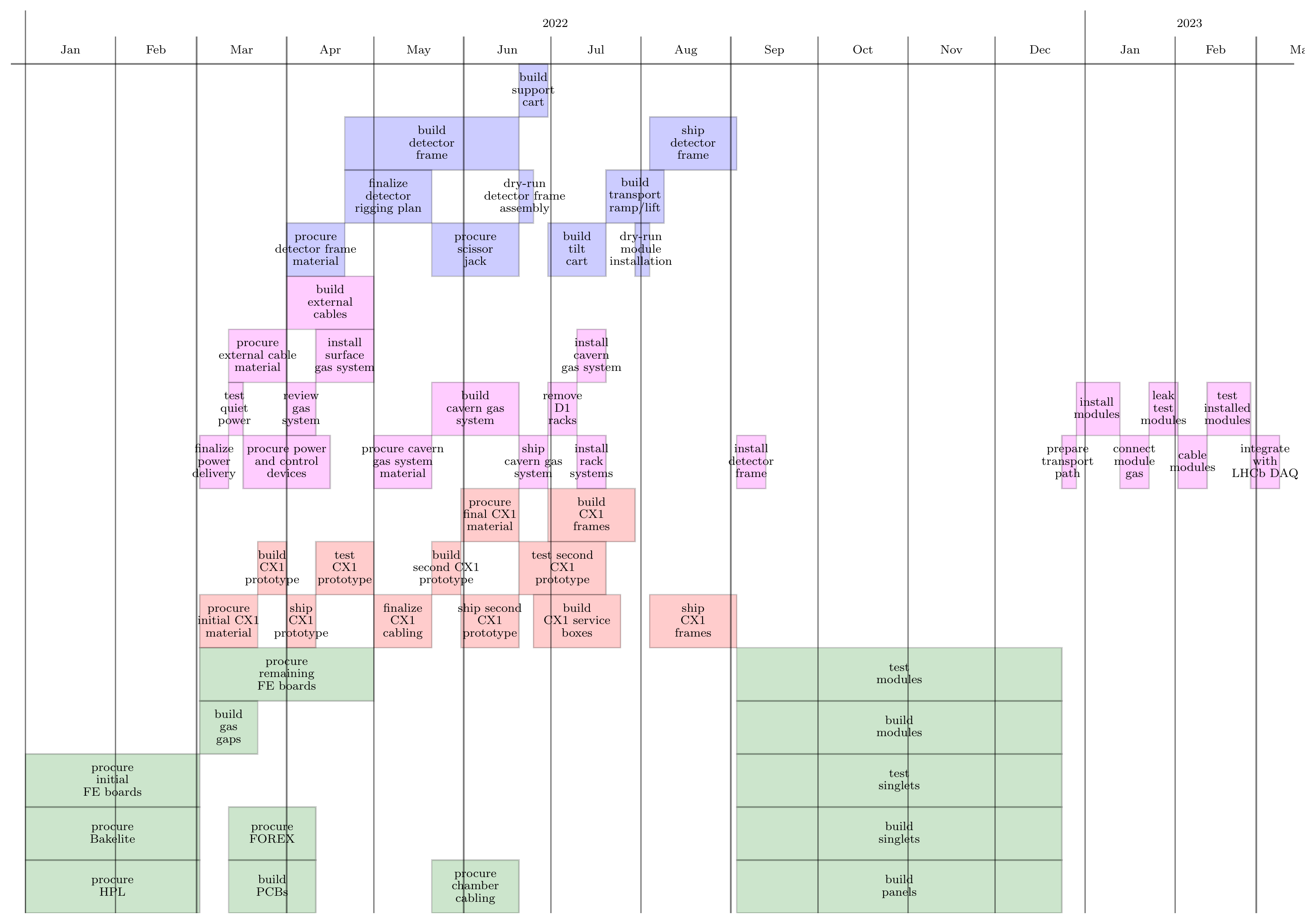}
    \caption{Initial timeline for the building and installation of \CODEXbeta. The tasks are split by work package: (blue) detector frame, (magenta) installation, (red) module frame, (green) RPCs.}
    \label{fig:timeline}
\end{figure*}

The initial timeline for the building and installation of \CODEXbeta is shown in Fig.~\ref{fig:timeline}.

\onecolumngrid
\phantom{The initial timeline for the building and installation of \CODEXbeta is shown in Fig.~\ref{fig:timeline}.}
\twocolumngrid

\input{whitepaper_v1.bbl}
\end{document}

%% file: authors_full.tex
\author{Giulio Aielli}
\affiliation{Universit\`a e INFN Sezione di Roma Tor Vergata, Roma, Italy}

\author{Juliette Alimena}
\affiliation{European Organization for Nuclear Research (CERN), Geneva, Switzerland}

\author{James Beacham}
\affiliation{Department of Physics, Duke University, Durham, NC 27706, United States}

\author{Eli Ben-Haim}
\affiliation{LPNHE, Sorbonne Universit{\'e}, Paris Diderot Sorbonne Paris Cit{\'e}, CNRS/IN2P3, Paris, France}

\author{Martino Borsato}
\affiliation{Kirchhoff-Institut f{\"u}r Physik (KIP), Ruprecht-Karls-Universit{\"a}t Heidelberg, Heidelberg, Germany}

\author{Matthew John Charles}
\affiliation{Universit\'e Pierre et Marie Curie, Paris, France}

\author{Xabier Cid Vidal}
\affiliation{Instituto Galego de F\'isica de Altas Enerx\'ias (IGFAE), Universidade de Santiago de Compostela, Santiago de Compostela, Spain}

\author{Victor Coco}
\affiliation{European Organization for Nuclear Research (CERN), Geneva, Switzerland}

\author{Albert De Roeck}
\affiliation{European Organization for Nuclear Research (CERN), Geneva, Switzerland}

\author{Biplab Dey}
\affiliation{ELTE E\"otv\"os Lor\'and University, Budapest, Hungary}

\author{Raphael Dumps}
\affiliation{European Organization for Nuclear Research (CERN), Geneva, Switzerland}

\author{Vladimir V.~Gligorov}
\affiliation{LPNHE, Sorbonne Universit{\'e}, Paris Diderot Sorbonne Paris Cit{\'e}, CNRS/IN2P3, Paris, France}
\affiliation{European Organization for Nuclear Research (CERN), Geneva, Switzerland}

\author{Rebeca Gonzalez Suarez}
\affiliation{Department of Physics and Astronomy, Uppsala Universitet, Uppsala, Sweden}

\author{Thomas Gorordo}
\affiliation{Physics Division, Lawrence Berkeley National Laboratory, Berkeley, CA 94720, USA}

\author{Louis Henry}
\affiliation{European Organization for Nuclear Research (CERN), Geneva, Switzerland}

\author{Philip Ilten}
\affiliation{Department of Physics, University of Cincinnati, Cincinnati, Ohio 45221, USA}

\author{Daniel Johnson}
\affiliation{Laboratory for Nuclear Science, Massachusetts Institute of Technology, Cambridge, MA 02139, USA}

\author{Simon Knapen}
\email{smknapen@lbl.gov}
\affiliation{Physics Division, Lawrence Berkeley National Laboratory, Berkeley, CA 94720, USA}
\affiliation{Department of Physics, University of California, Berkeley, CA 94720, USA}

\author{Olivier Le Dortz}
\affiliation{LPNHE, Sorbonne Universit{\'e}, Paris Diderot Sorbonne Paris Cit{\'e}, CNRS/IN2P3, Paris, France}

\author{Saul L\'{o}pez Soli\~{n}o}
\affiliation{Instituto Galego de F\'isica de Altas Enerx\'ias (IGFAE), Universidade de Santiago de Compostela, Santiago de Compostela, Spain}

\author{Titus Mombächer}
\affiliation{Instituto Galego de F\'isica de Altas Enerx\'ias (IGFAE), Universidade de Santiago de Compostela, Santiago de Compostela, Spain}

\author{Benjamin Nachman}
\affiliation{Physics Division, Lawrence Berkeley National Laboratory, Berkeley, CA 94720, USA}

\author{David T. Northacker}
\affiliation{Department of Physics, University of Cincinnati, Cincinnati, Ohio 45221, USA}

\author{Michele Papucci}
\affiliation{Walter Burke Institute for Theoretical Physics, California Institute of Technology, Pasadena, CA 91125, USA}

\author{Gabriella P\'{a}sztor}
\affiliation{ELTE E\"otv\"os Lor\'and University, Budapest, Hungary}

\author{Luca Pizzimento}
\affiliation{Dipartimento di Fisica, Universit{\`a} degli Studi di Roma ``Tor Vergata'', Rome, Italy}

\author{Francesco Polci}
\affiliation{LPNHE, Sorbonne Universit{\'e}, Paris Diderot Sorbonne Paris Cit{\'e}, CNRS/IN2P3, Paris, France}

\author{Dean J.~Robinson}
\email{drobinson@lbl.gov}
\affiliation{Physics Division, Lawrence Berkeley National Laboratory, Berkeley, CA 94720, USA}
\affiliation{Department of Physics, University of California, Berkeley, CA 94720, USA}

\author{Heinrich Schindler}
\affiliation{European Organization for Nuclear Research (CERN), Geneva, Switzerland}

\author{Michael D. Sokoloff}
\affiliation{Department of Physics, University of Cincinnati, Cincinnati, Ohio 45221, USA}

\author{Aditya Suresh}
\affiliation{Physics Division, Lawrence Berkeley National Laboratory, Berkeley, CA 94720, USA}
\affiliation{Department of Physics, University of California, Berkeley, CA 94720, USA}

\author{Paul Swallow}
\affiliation{University of Birmingham, Birmingham, United Kingdom}

\author{Riccardo Vari}
\affiliation{INFN Sezione di Roma La Sapienza, Roma, Italy}

\author{G\'{a}bor Veres}
\affiliation{ELTE E\"otv\"os Lor\'and University, Budapest, Hungary}

\author{Carlos V\'azquez~Sierra}
\affiliation{European Organization for Nuclear Research (CERN), Geneva, Switzerland}

\author{Nigel Watson}
\affiliation{University of Birmingham, Birmingham, United Kingdom}

\author{Michael K. Wilkinson}
\affiliation{Department of Physics, University of Cincinnati, Cincinnati, Ohio 45221, USA}

\author{Michael Williams}
\affiliation{Laboratory for Nuclear Science, Massachusetts Institute of Technology, Cambridge, MA 02139, USA}

\author{Emilio Xos\'e Rodr\'iguez Fern\'andez}
\affiliation{Instituto Galego de F\'isica de Altas Enerx\'ias (IGFAE), Universidade de Santiago de Compostela, Santiago de Compostela, Spain}